\documentclass[pdflatex,iicol,sn-mathphys-num]{sn-jnl}

\usepackage{graphicx}%
\usepackage{amsmath,amssymb,amsfonts}%

\usepackage[version=4]{mhchem}
\usepackage{ifthen}

\usepackage{siunitx}
\DeclareSIUnit{\Molar}{\textsc{M}}
\DeclareSIUnit{\molar}{\textsc{M}}
\DeclareSIUnit{\mmolgh}{\mmol\per\gram\per\hour}

\def\mode{clean}            

\ifthenelse{\equal{\mode}{reply}}{
    \def\changesmode{draft}
}{
    \ifthenelse{\equal{\mode}{newchanges}}{
        \def\changesmode{final}
    }{
        \def\changesmode{final}
    }
}
\usepackage[\changesmode]{changes}
\ifthenelse{\equal{\changesmode}{draft}}{
    \usepackage{soul}
    \usepackage[normalem]{ulem}
    \renewcommand{\st}[1]{\relax\ifmmode\text{\sout{\ensuremath{#1}}}\else\sout{#1}\fi}
    \setaddedmarkup{\textcolor{red}{#1}}
    \setdeletedmarkup{\textcolor{black}{\st{#1}}}
    \let\origadded\added
    \let\origdeleted\deleted
    \let\origreplaced\replaced
    \let\origlistofchanges\listofchanges
    \renewcommand{\added}[1]{\relax\ifmmode\origadded{\ensuremath{#1}}
                                   \else\origadded{#1}\fi}
    \renewcommand{\deleted}[1]{\relax\ifmmode\origdeleted{\ensuremath{#1}}
                                     \else\origdeleted{#1}\fi}
    \renewcommand{\replaced}[2]{\relax\ifmmode{\origreplaced{\ensuremath{#1}}
                                                            {\ensuremath{#2}}}
                                      \else{\origreplaced{#1}{#2}}\fi}
    \renewcommand\listofchanges{\newpage\onecolumngrid\origlistofchanges\newpage}
    \newcommand{\newadd}[1]{}
}{
    \ifthenelse{\equal{\mode}{newchanges}}{
        \newcommand{\newadd}[1]{\textcolor{red}{#1}}
    }{
        \newcommand{\newadd}[1]{#1}
    }
}
\makeatletter
    \let\origsum\sum@
\makeatother

\newcommand{\mysub}[1]{\textnormal{\textsc{#1}}}
\newcommand{\lact}{{\mysub{l}}}
\newcommand{\gluc}{{\mysub{g}}}
\newcommand{\ox}{{\mysub{o}}}
\newcommand{\atp}{\mysub{atp}}
\newcommand{\maint}{\mysub{m}}

\newcommand{\expect}[1]{\left\langle #1 \right\rangle}
\newcommand{\avgul}{\overline{u_\lact\vphantom{+}}}
\newcommand{\supnote}[1]{Supplementary Note~#1}
\newcommand{\supmethod}[1]{Supplementary Method~#1}
\newcommand{\suptable}[1]{Supplementary Table~#1}

\begin{document}

\title[Metabolic coordination and phase transitions]{Metabolic coordination and phase transitions in spatially distributed multi-cellular systems}


\author[1,2]{\fnm{Krishnadev} \sur{Narayanankutty}}

\author[3,4]{\fnm{Jos\'e Antonio} \sur{Pereiro-Morejon}}

\author[1,2]{\fnm{Ari\'an} \sur{Ferrero-Fern\'andez}}

\author[5]{\fnm{Valentina} \sur{Onesto}}

\author[5]{\fnm{Stefania} \sur{Forciniti}}

\author[5]{\fnm{Loretta} \sur{L. del Mercato}}

\author[6]{\fnm{Roberto} \sur{Mulet}}

\author*[7,8]{\fnm{Andrea} \sur{De Martino}} \email{andrea.demartino@polito.it}

\author*[9]{\fnm{David} \sur{S. Tourigny}} \email{d.tourigny@bham.ac.uk}

\author*[1,10]{\fnm{Daniele} \sur{De Martino}} \email{daniele.demartino@ehu.eus}

\affil[1]{\orgname{Biofisika Institute (CSIC, UPV/EHU)}, \orgaddress{\street{Barrio Sarriena s/n}, \city{Leioa}, \postcode{48940}, \state{Bizkaia}, \country{Spain}}}

\affil[2]{\orgdiv{Department of Molecular Biology and Biochemistry}, \orgname{University of the Basque Country}, \orgaddress{\city{Leioa}, \postcode{48940}, \state{Bizkaia}, \country{Spain}}}

\affil[3]{\orgdiv{Group of Complex Systems and Statistical Physics, Physics Faculty}, \orgname{University of Havana}, \orgaddress{\street{San Lazaro y L, Vedado}, \city{La Habana}, \postcode{10400}, \state{La Habana}, \country{Cuba}}}

\affil[4]{\orgdiv{Biology Faculty}, \orgname{University of Havana}, \orgaddress{\street{San Lazaro y L, Vedado}, \city{La Habana}, \postcode{10400}, \state{La Habana}, \country{Cuba}}}

\affil[5]{\orgname{Institute of Nanotechnology, National Research Council (CNR-NANOTEC)}, \orgaddress{\street{Campus Ecotekne, via Monteroni}, \city{Lecce}, \postcode{73100}, \state{Puglia}, \country{Italy}}}

\affil[6]{\orgdiv{Department of Theoretical Physics}, \orgname{University of Havana}, \orgaddress{\street{San Lazaro y L, Vedado}, \city{La Habana}, \postcode{10400}, \state{La Habana}, \country{Cuba}}}

\affil[7]{\orgname{Politecnico di Torino}, \orgaddress{\street{Corso Duca degli Abruzzi 24}, \city{Torino} \postcode{10129}, \country{Italy}}}

\affil[8]{\orgname{Italian Institute for Genomic Medicine, c/o IRCCS}, \orgaddress{\street{SP 142 Km 3.95}, \city{Candiolo} \postcode{10060}, \country{Italy}}}

\affil[9]{\orgdiv{School of Mathematics}, \orgname{University of Birmingham}, \orgaddress{\street{Edgbaston}, \city{Birmingham}, \postcode{B15 2TT}, \country{United Kingdom}}}

\affil[10]{\orgname{Ikerbasque -- Basque Foundation for Science}, \orgaddress{\street{Plaza Euskadi 5}, \city{Bilbao}, \postcode{48009}, \state{Bizkaia}, \country{Spain}}}

\abstract{During overflow metabolism, cells excrete glycolytic byproducts when growing under aerobic conditions in a seemingly wasteful fashion. While potentially advantageous for microbes with finite oxidative capacity, its role in higher organisms is harder to assess. Recent single-cell experiments suggest overflow metabolism arises due to imbalances in inter-cellular exchange networks. We quantitatively characterize this scenario by integrating spatial metabolic modeling with tools from statistical physics and experimental single-cell flux data. Our results provide a theoretical demonstration of how diffusion-limited exchanges shape the space of accessible multi-cellular metabolic states. Specifically, a phase transition from a balanced network of exchanges to an unbalanced, overflow regime occurs as mean glucose and oxygen uptake rates vary. Heterogeneous single-cell metabolic phenotypes occur near this transition. Time-resolved tumor-stroma co-culture data support the idea that overflow metabolism stems from failure of inter-cellular metabolic coordination. In summary, environmental control is an emergent multi-cellular property, rather than a cell-autonomous effect.}




\maketitle

\section{Introduction}\label{introduction}

Cell populations adapt to an environment on at least two different levels: (i) via intra-cellular regulation (e.g. metabolic, signaling, genetic), which underlies essential maintenance, biosynthetic and, possibly replicative processes; and (ii) via extra-cellular mechanisms (e.g. sensing, signaling, motility), necessary to harvest information and control exchanges with the medium.
The latter level prompts inter-cellular interactions and introduces an ecological dimension to multi-cellular systems, where cells can be seen as agents that need to meet certain requirements while jointly modulating a shared environment. The way in which the two layers integrate is a key determinant of  adaptation, viability, and ultimately fitness \cite{Pfeiffer2001,Vaupel2010,Waltermann2011,Caulin2011}.
This raises a rather basic question: is a viable environment the result of the straightforward aggregation of a large number of autonomous actions by individual cells, or is it rather an emergent property of the collective behavior of many interacting cells? \added{In the former case, intra-cellular  constraints (and possibly cell-specific objective functions) effectively direct population behavior \cite{de2020common}. In the latter, inter-cellular interactions play the central role \cite{damiani2011cell}. Can one quantitatively separate the two contributions?}

The contours of this problem become especially clear upon focusing on a cellular process that directly links the two levels described above, namely carbon overflow (CO). In short, CO consists of the excretion into the medium of carbon-based waste products of intra-cellular carbon catabolism, such as acetate, ethanol, or lactate, in aerobic conditions \cite{wolfe2005acetate, vazquez2017overflow}.
This may occur \replaced{for instance because}{when} the cell's oxidative capacity is saturated (e.g. due to excess glucose availability or dysregulated import pathways), so that, even in the presence of abundant oxygen, incoming carbon is diverted towards  fermentation \cite{de2020common, wang2022saturation, Vazquez2008, Shlomi2011}.
CO appears to be a `universal' feature of cellular metabolism, which has been consistently observed across domains. While in microbial systems it can be partly explained by the evolutionary advantage of a higher growth rate at the cost of lower energy yield \cite{basan2015overflow, szenk2017fast, mori2019yield, chen2019energy}, the root cause of CO in higher organisms or in tumors (where it underlies the Warburg effect \cite{vander2009understanding}) is more difficult to ascribe, especially when it is not associated with growth and replication \cite{Caulin2011, Hirschhaeuser2011, Vazquez2016a, Sullivan2016, kukurugya2022warburg, shen2024mitochondrial}.

An ecological role for CO derives from the fact that accumulation of fermentation byproducts in the extra-cellular medium leads to acidification and, in turn, detrimental effects ranging from the slowdown of protein synthesis to growth inhibition due to apoptosis \cite{de2019lactate}. Remediation of the shared environment is therefore of paramount importance for cell populations.
On the other hand, exporters of overflow products effectively supply the medium with additional carbon equivalents on which cells can rely for sustenance (via oxidation)  \cite{mason2017lactate,millard2022toxic}. Importers thereby contribute to correcting environmental pollution. Taken together, these processes are major drivers of functionally- and ecologically-significant inter-cellular crosstalk \cite{liberti2016warburg,fernandez2017microenvironmental,wang2022lactate,hershey2023clonal}.
Nevertheless, how decisive such a crosstalk is compared to individual cellular decisions in engineering the environment remains an open question.

We attempt to address this issue by integrating (a) constraint-based metabolic modeling (CBM) \cite{bordbar2014constraint} and diffusion constraints, (b) the statistical physics approach to the study of emergent \deleted{and collective} phenomena \cite{nishimori2011elements}, and (c) high-resolution data for single-cell behavior in a cell population adapting to a glucose-based medium \cite{onesto2022probing}.
More specifically, we first employ CBM to demonstrate that a spatially distributed population of cells with identical metabolic requirements coupled through an exchange network undergoes a crossover between two distinct metabolic regimes when the population-averaged metabolic rate changes. The first regime is characterized by weak medium acidification, while significant accumulation of overflow products occurs in the second.
Next, we rationalize these findings mathematically using \replaced{a highly stylized but}{an} analytically tractable \replaced{version}{approximation} of the  model. Within this approach, the crossover takes the form of an order-disorder transition similar to those that characterize thermodynamic systems in statistical physics \cite{nishimori2011elements}, and bulk overflow appears as an emergent feature of a population of interacting cells.
Finally, \added{building on the theory, we exploit} statistical inference techniques \added{to reconstruct the metabolic trajectory of} an experimentally-studied mixed population of cancer cells and cancer-associated fibroblasts \added{that displays the Warburg effect and} self-organizes its collective metabolism over time \replaced{to reduce lactate spillover via metabolic exchanges.
\newadd{We will focus on the lag phase with negligible glucose depletion and growth, because it provides insights into the cellular and molecular mechanisms that govern cancer cell adaptation and subsequent proliferation.}
We show in particular that the population, while coordinating at metabolic level and reducing lactate spillover, collectively moves towards states where the ATP yield on glucose is optimal. This provides a quantitative, low-dimensional, and interpretable representation of the complex, high-dimensional dynamics of the population in the space of feasible metabolic states.}{so that it remains close to the theoretical critical line separating the two regimes, i.e. at the onset of acidification where metabolic heterogeneity is large.}

\section{Results}\label{results}

\subsection{Single-cell metabolic model and flux space}

\begin{figure*}[htp]
    \centering
    \includegraphics[width=\textwidth]{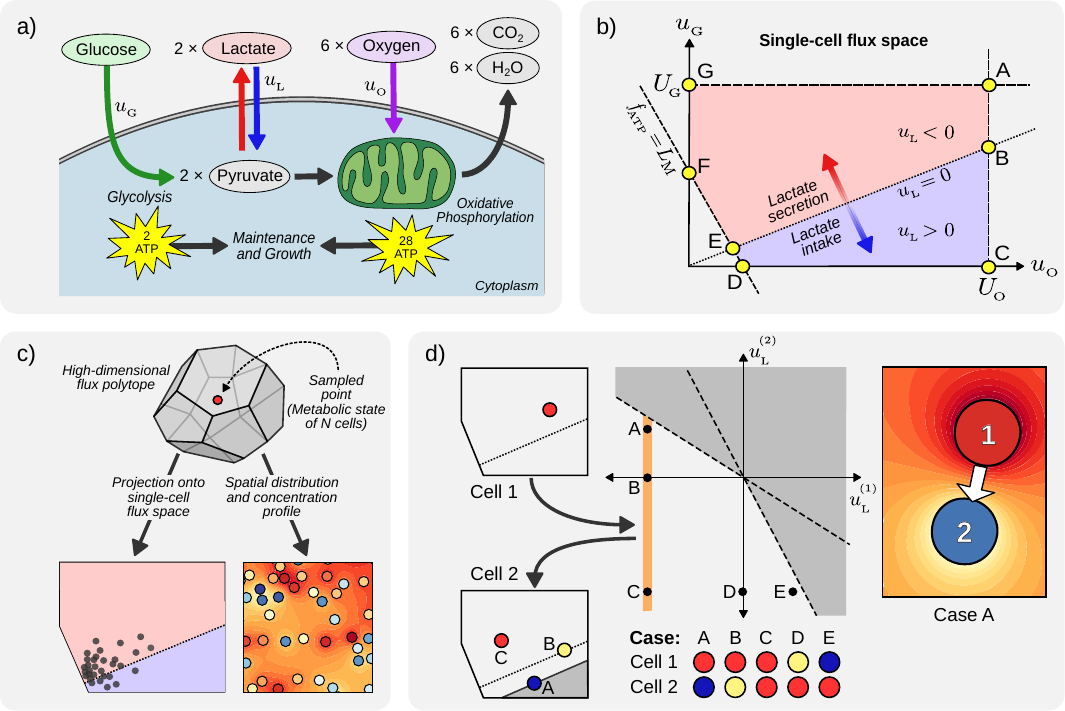}
    \caption{Constraint-based metabolic model.
    (a) Sketch of the single-cell metabolic network representing the central carbon pathways ($u_\gluc$ glucose uptake, $u_\lact$ lactate flux, $u_\ox$ oxygen uptake).
    (b) Feasible single-cell flux space ($\mathcal{F}_1$) in the $(u_\ox,u_\gluc)$ plane, bounded by Eqs (\ref{eq:u_gluc}) (glucose intake), (\ref{eq:u_ox}) (oxygen intake), and (\ref{eq:L_M}) (ATP maintenance).  For a single cell (as well as for the average bulk behavior) the purple region ($u_\lact>0$ or lactate import) is unfeasible unless lactate is exogenously provided. Points \textsf{A-G} are those where the cell would maximize rate of ATP production (\textsf{A}), maximize rate of ATP production with zero net lactate exchange (\textsf{B}), maximize rate of ATP production while using lactate as the only carbon source (\textsf{C}), minimize rate of ATP production while using lactate as the only carbon source (\textsf{D}), minimize rate of ATP production with zero net lactate exchange (\textsf{E}), minimize rate of ATP production anaerobically (\textsf{F}), or maximize rate of lactate excretion (\textsf{G}).
    (c) The possibility for cells to exchange lactate defines an extended  metabolic flux space for a system of $N$ cells ($\mathcal{F}_N$) whose configurations can be projected into single-cell flux space. In turn, the corresponding lactate fluxes define a spatial concentration gradient in the medium via (\ref{eq:profile}).
    (d) Case $N=2$, feasible space in the plane $(u_{\lact}^{(1)},u_\lact^{(2)})$. In cases A and E cells are coupled via lactate exchange, one cell acting as a donor (red) for the other (blue). This makes the $u_\lact>0$ (purple) region of panel (B) viable for the acceptor cell even in absence of an external lactate source.
    }
\label{fig:network}
\end{figure*}

To build our theoretical setup, we begin by modeling the metabolic flux space available to a single mammalian cell through CBM with the minimal reaction network for energy production by central carbon pathways displayed in Fig.~\ref{fig:network}a. \added{This simplistic model consists of just three metabolic reactions that carry out (a) the import and conversion of glucose to pyruvate (representing glycolysis), (b) the interconversion of pyruvate and lactate (representing lactate import/export), and (c) the generation of ATP using pyruvate and oxygen (representing oxidative phosphorylation).} We want to model a typical  overflow scenario like the one presented in \cite{onesto2022probing}, where the medium is glucose enriched, there is lactate accumulation in aerobic conditions and metabolism is running mainly  for the sake of energy production with negligible biomass build-up. \added{This corresponds to the initial part of the growth curve, during which the population adapts to the environment.}

\added{Because the experimental timescales we consider (hours after seeding) are much longer than typical metabolite turnover times (seconds for ATP) \cite{milo2015cell, park2016metabolite, de2016genome}, we can make the same stationarity assumption for metabolite levels that is standard in approaches based on Flux Balance Analysis \cite{orth2010flux}}. At steady state, the equation for carbon mass-balance is
\begin{equation}
    u_\gluc  +\frac{u_\lact}{2} - \frac{u_\ox}{6} = 0~~,
\label{eq:mass_balance}
\end{equation}
where $u_\gluc\geq 0$ is the flux for import of glucose from the environment (in units
\si{\mmolgh}),
$u_\lact$ is the corresponding lactate exchange flux ($u_\lact>0$ for import, $u_\lact<0$ for export) and $u_\ox\geq 0$ is the flux for import of molecular oxygen (that we assume corresponds to respiratory flux, so that it enters the carbon mass-balance due to its equivalence to carbon dioxide secretion).
The net rate of ATP production is \added{instead} given by
\begin{equation}\label{eq:fatp}
    f_{\atp} = -u_\lact + 5 u_\ox ~~,
\end{equation}
where we used empirical coefficients for ATP produced by respiration and fermentation pathways (\supnote{1}). Along with (\ref{eq:mass_balance}), the space of viable fluxes is defined by the additional constraints
\begin{gather}
    0 \leq u_\gluc \leq U_\gluc         \label{eq:u_gluc}\\
    0 \leq u_\ox \leq U_\ox         \label{eq:u_ox}\\
    f_{\atp} \geq L_\maint   \label{eq:L_M}
\end{gather}
representing respectively the limited capacities of glucose import channels (Eq. (\ref{eq:u_gluc})) and mitochondrial activity (Eq. (\ref{eq:u_ox})), and a minimal rate of energy production required for cell maintenance (Eq. (\ref{eq:L_M})). Values for the parameters $U_\gluc$, $U_\ox$ and $L_\maint$ are also available from the literature (\supnote{1}).

The resulting single-cell flux space is represented by the 2-dimensional polytope in the $(u_\ox,u_\gluc)$ plane shown in Fig.~\ref{fig:network}b, whose oblique boundary corresponds to the maintenance requirement (\ref{eq:L_M}). Lactate fluxes can be read off from (\ref{eq:mass_balance}).
The line separating the purple and pink shaded regions corresponds to states with $u_\lact=0$. States with constant but non-zero lactate exchange are represented by lines parallel to it. In particular, states with $u_\lact>0$ (resp. $u_\lact<0$), i.e. with lactate import (resp. export) foliate the purple (resp. pink) shaded region.
In the absence of environmental lactate, an isolated cell can only attain states with $u_\lact\leq 0$ (\deleted{no} lactate import \added{is not allowed}). In the presence of a lactate source, however, states with $u_\lact>0$ become accessible (lactate import \added{is allowed}).
We shall use the symbol $\mathcal{F}_1$ to denote the viable single-cell flux space of Fig.~\ref{fig:network}b. An independent cell that autonomously adjusts its metabolism to optimize a linear objective function (as in Flux Balance Analysis \cite{orth2010flux}) would be found at specific points on the boundary of $\mathcal{F}_1$.
For instance (Fig.~\ref{fig:network}b), at point \textsf{A} it maximizes the rate of ATP production. (More examples are given in the caption of Fig.~\ref{fig:network}.)

\subsection{Exchange coupling and multi-cellular flux space}

When $N$ cells share the same extra-cellular environment, lactate-excreting cells effectively act as lactate sources. This makes the $u_\lact>0$ portion of $\mathcal{F}_1$ (purple region in Fig.~\ref{fig:network}b) potentially accessible to other cells even if there is no external source of lactate in the culture.
Cells therefore become metabolically coupled through the exchange of lactate, as endogenous lactate is shuttled across the population by diffusion.
By modeling cells as spherical sources or sinks of lactate, with radius $R$ and located at fixed positions $\mathbf{r}_i$ ($i=1,2,...,N$), one can show that, at steady state
\added{(i.e. in practice for timescales larger than the diffusion time of lactate across the experimental length-scale we consider, which roughly equals $L^2/D_\lact \simeq 4$ min, with $L=0.5$ mm the system size and  $D_\lact\simeq \qty{700}{\micro\meter\squared\per\second}$ the diffusion constant of lactate)}, the lactate exchange fluxes $u_\lact^{(i)}$ of all cells must obey an additional set of $N$ constraints described by (\supnote{2}; see also \cite{capuani2015quantitative})
\begin{gather}
\label{eq:diffusion}
    \origsum_{j=1}^N A_{ij} u^{(j)}_\lact  \leq 0  \qquad (i = 1,2,\ldots,N)~~,\\
\label{eq:Aij}
    A_{ij}
    = \delta_{ij} + (1-\delta_{ij}) \frac{R}{|\mathbf{r}_i-\mathbf{r}_j|}>0 ~~,
\end{gather}
where $|\mathbf{r}_i-\mathbf{r}_j|\geq 2R$  and $\delta_{ij}=1$ if $i=j$ and zero otherwise.
\added{In rough words, these constraints effectively couple cells by imposing that the net consumption of lactate across the culture can not exceed its endogenous supply.
However, the generalization that accounts for an exogenous lactate source or for lactate accumulated in the culture is straightforward (\supmethod{4}).  In principle, similar diffusional coupling constraints hold for glucose and oxygen. For the cell densities we consider, though, they are immaterial (i.e. cells do not compete for either of the two, see \supmethod{5}).}
Combined with $N$ copies of \deleted{constraints} (\ref{eq:mass_balance}) and (\ref{eq:u_gluc})-(\ref{eq:L_M}), one for each cell $i$, the inequalities (\ref{eq:diffusion}) define a $2N$-dimensional convex polytope containing the feasible flux configurations of a system of $N$ cells coupled through diffusion-limited lactate exchanges (Fig.~\ref{fig:network}c).
We henceforth denote this \deleted{total} space by $\mathcal{F}_N$. In turn, each point in $\mathcal{F}_N$ can be represented by $N$ points in the single-cell space $\mathcal{F}_1$ (one per cell, Fig.~\ref{fig:network}c).
Finally, it is possible to reconstruct the spatial concentration profile of lactate by assuming that lactate levels in the culture obey the Laplace equation. Specifically, the concentration at position $\mathbf{r}$ is given by (\supnote{2})
\begin{equation}\label{eq:profile}
    c_\lact(\mathbf{r}) = \origsum_{i=1}^N
                        \frac{u_\lact^{(i)}}{D_\lact|\mathbf{r}
                        -   \mathbf{r}_i|}+B(\mathbf{r})~~,
\end{equation}
where $B(\mathbf{r})$ is a term accounting for exchanges with the boundary of the system. (Note that we are assuming that cells are identical, in that the parameters $U_\gluc, U_\ox, L_\maint$ and $R$ are the same for each cell.) \deleted{, and that there is no competition for nutrients other than lactate.)}

The effect of the diffusion constraint is most easily visualized diagrammatically in the case of two cells ($N=2$, Fig.~\ref{fig:network}d), where the maximal lactate uptake rate for one cell is determined by both a minimum rate of lactate production by the other and the distance between them (reflected in the slopes of the dashed lines in Fig.~\ref{fig:network}d).
Clearly, (\ref{eq:diffusion}) does not allow for both cells to import lactate ($u_\lact^{(1)}>0$, $u_\lact^{(2)}>0$) if none is supplied externally. This picture generalizes to $N$ cells: it can be shown (\supnote{3}) that environmental lactate accumulation ($\origsum_{i=1}^N u_\lact^{(i)}$) can not be zero unless the lactate flux of each individual cell vanishes.
Conversely, if lactate is exchanged between any cells within the population then there must be some non-zero leakage of lactate in the medium, i.e. an accumulation of overflow product akin to the Warburg effect.

\begin{figure*}[htp]
    \centering
    \includegraphics[width=\textwidth]{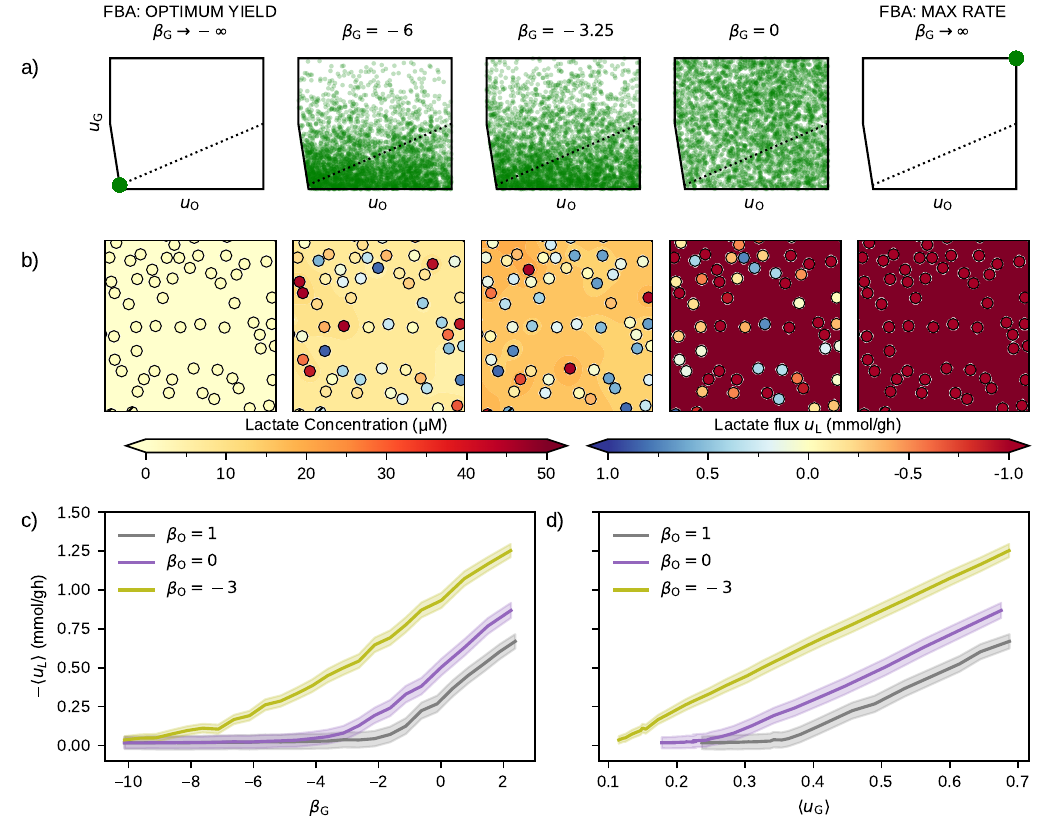}
    \caption{Emergent bulk overflow in CBM simulations. Simulations are performed by sampling the feasible space $\mathcal{F}_N$ for $N=150$ cells distributed over an area of $500 \times \qty{500}{\micro\meter\squared}$ according to (\ref{eq:boltz2}) via Hit-and-Run Monte Carlo Markov chains for fixed $\beta_\ox=0$ and different values of $\beta_\gluc$. This interpolates between the states of maximum ATP yield ($\beta_\gluc \to -\infty$) and  maximum ATP rate ($\beta_\gluc \to \infty$).
    (a) Multi-cellular flux configurations projected into single-cell space in the $(u_\ox,u_\gluc)$ plane (same as in Fig.~\ref{fig:network}b) for different values of $\beta_\gluc$.
    (b) Typical spatial lactate concentration gradient across the culture (background color) and single cell lactate fluxes (colored circles) for the same values of $\beta_\gluc$ as in (a).
    (c) Mean net lactate excretion fluxes ($-\expect{u_\lact}$) as a function of $\beta_\gluc$ and (d) of the mean glucose uptake for different values of $\beta_\ox$. The shaded region indicates standard error on the mean. Notice the approximately threshold-linear behavior for larger values of $\beta_\ox$.}
    \label{fig:sampling}
\end{figure*}

\subsection{Environmental lactate spillover as a failure of inter-cellular coordination}

While constraints (\ref{eq:mass_balance}) and (\ref{eq:u_gluc})-(\ref{eq:diffusion}) define the space $\mathcal{F}_N$ of feasible multi-cellular metabolic flux configurations and implicitly include a lactate exchange network, \replaced{interpreting $\mathcal{F}_N$}{an interpretation} in terms of population-level metabolic states is not straightforward.
To gain a deeper understanding, we resorted to a statistical approach based on sampling $\mathcal{F}_N$ according to a controllable probabilistic rule. A simple and theoretically convenient choice for a probability density over the $2N$-dimensional flux space is given by the Boltzmann distribution \cite{de2018introduction}
\begin{equation}
\label{eq:boltz}
    p(\mathbf{\underline{u}}|\beta)
        =\frac{\exp[\beta h(\mathbf{\underline{u}})]}{Z(\beta)}
        ~~~~~~
        (\mathbf{\underline{u}}\in\mathcal{F}_N)~~,
\end{equation}
where $\mathbf{\underline{u}}\equiv\{(u_\gluc^{(i)},u_\ox^{(i)})\}_{i=1}^N$ denotes an $N$-cell configuration of metabolic degrees of freedom, $\beta$ is a numerical parameter, $h$ is a prescribed function of the glucose and oxygen import fluxes of every cell, while
\begin{equation}
    Z(\beta)    =   \int_{\mathcal{F}_N}
                        \exp[\beta h(\mathbf{\underline{u}})]
                            d\mathbf{\underline{u}}
\end{equation}
is a factor ensuring normalization over $\mathcal{F}_N$. For $\beta=0$, (\ref{eq:boltz}) coincides with the uniform distribution over $\mathcal{F}_N$, under which each viable $N$-cell state is equally likely. When $\beta\to+\infty$ (resp. $\beta\to-\infty$), instead, the sampling concentrates on $N$-cell states of maximum (resp. minimum) $h$. Notably, since
\begin{gather}\label{eq:avg_h}
    \expect{h}_\beta
    \equiv
    \frac{1}{N} \int_{\mathcal{F}_N}
                    h(\mathbf{\underline{u}})
                    p(\mathbf{\added{\underline{u}}}|\beta)
                    d\mathbf{\underline{u}}
    =
    \frac{\partial}{\partial\beta}\ln Z(\beta)~~,\\
    \sigma^2_{h}
    \equiv
    \expect{h^2}_\beta - \expect{h}_\beta^2
    =  \frac{\partial^2}{\partial\beta^2} \ln Z(\beta)~~,
\label{eq:var_h}
\end{gather}
fixing the value of $\beta$ is equivalent to constraining the population-averaged value of $h$ while still allowing for variability in single-cell metabolic profiles. Most importantly, for any $\beta$, (\ref{eq:boltz}) describes the distribution with constrained mean value of $h$ having {\it maximum entropy} and, therefore, \replaced{minimum extra bias}{thus being least unbiased} \cite{de2018introduction}. As $\beta$ changes, therefore, (\ref{eq:boltz}) allows \added{for} the exploration of a broad range of population-level features.

\replaced{Previous applications of  maximum-entropy frameworks to metabolic data mainly focused on bacterial growth, where the biomass synthesis rate is a natural choice for $h$ \cite{de2016growth,de2018statistical}.
Here we want to focus instead on the adaptation part of the dynamics, during which biomass synthesis is negligible  \cite{onesto2022probing}. In this regime, natural choices for $h$ are the overall glucose and oxygen import fluxes, as they represent the key independent degrees of freedom the culture can coordinate to control. Any other linear function (e.g. ATP production) can be obtained from these via linear combination. We}{In our specific case, the overall glucose and oxygen import fluxes represent natural choices for $h$. To gain
additional sampling freedom, we} therefore opted for a version of (\ref{eq:boltz}) that allows to maximize or minimize these quantities independently, i.e. ($\mathbf{\underline{u}}\in\mathcal{F}_N$)
\begin{equation}
\label{eq:boltz2}
    p(\mathbf{\underline{u}}|\beta_\gluc,\beta_\ox)
        \propto
        \exp    \left[\beta_\gluc\origsum_{i=1}^N u_\gluc^{(i)}
                        + \beta_\ox \origsum_{i=1}^N u_\ox^{(i)}    \right]~~,
\end{equation}
where $\beta_\gluc$ and $\beta_\ox$ are real parameters, \added{and sampled} the space $\mathcal{F}_N$ \deleted{has been then sampled} according to (\ref{eq:boltz2}) for different values of $\beta_\gluc$ and $\beta_\ox$ via Hit-and-Run Monte Carlo (\supmethod{2}, and \cite{de2018statistical}).
(Identities similar to (\ref{eq:avg_h}) and (\ref{eq:var_h}) valid for (\ref{eq:boltz2}) are given in \supmethod{1}). \deleted{Key} Representative configurations for $150$ cells uniformly  scattered at random
over an area of $0.5 \times 0.5$ mm$^2$  are showcased in Fig.~\ref{fig:sampling}a.

We focus \deleted{our attention} on the case where $\beta_\ox$ is fixed to a finite value and $\beta_\gluc$ is varied from $-\infty$ to $+\infty$ (Fig.~\ref{fig:sampling}a and b), thereby modulating the mean glucose consumption in the population \added{from minimal to maximal}.
When $\beta_\gluc\to-\infty$, cells  independently maximize their ATP/glucose yields (Fig.~\ref{fig:sampling}a, leftmost panel), corresponding to point \textsf{E} in Fig.~\ref{fig:network}b.
With our (biologically plausible) choice of parameters, this leads to a homogeneous configuration where all cells run on respiration, none produce lactate, and thus no lactate accumulates in the medium (Fig.~\ref{fig:sampling}b, leftmost panel). As $\beta_\gluc$ increases, cells begin to excrete and import lactate, leading to highly heterogeneous lactate fluxes.
Remarkably, the resulting spillover is very small, implying that the lactate exchange network is nearly balanced in spite of the presence of large excretion fluxes.
However, a further increase in $\beta_\gluc$ destabilizes the network, causing lactate to accumulate in the medium at significant levels  (Fig.~\ref{fig:sampling}a--b, middle panels).
Finally, for $\beta_\gluc\to+\infty$, when cells independently maximize their ATP production rates (Fig.~\ref{fig:sampling}a, rightmost panel), the population returns to a homogeneous state in which all cells run strongly on fermentation and excrete large amounts of lactate (Fig.~\ref{fig:sampling}b, rightmost panel). This corresponds to point \textsf{A} in Fig.~\ref{fig:network}b.

The crossover from the state of maximum ATP yield to the state of maximum ATP rate is clearly reflected by the fact that the average net flux of lactate excretion across the population ($-\expect{u_\lact}$) increases as $\beta_\gluc$ increases (Fig.~\ref{fig:sampling}c).
We observe that a range of values of $\beta_\ox$ exists for which $-\expect{u_\lact}$ closely resembles a threshold-linear function of $\beta_\gluc$, marking a sharp transition between regimes with small and large lactate spillover, respectively.
Such a behavior is reminiscent of that of an order parameter in standard order-disorder transitions in statistical physics (see e.g. \cite{parisi1998statistical}, Ch.~3).

To summarize, numerical exploration of the feasible space of a multi-cellular  metabolic system subject to diffusion-limited lactate exchange suggests that the metabolic activity of cells gives rise to a complex and highly heterogeneous interaction network that couples lactate producers to lactate importers.
Environmental lactate accumulation may emerge from imbalances in this network that, depending on the mean oxygen consumption ($\beta_\ox$), can set in abruptly as $\beta_\gluc$ (i.e. the average net glucose consumption rate) increases. \added{Finally,} the model is able to reproduce in a stylized way typical bulk overflow-rate curves \cite{vazquez2017overflow, basan2015overflow, niebel2019upper}, unraveling the underlying single cell dynamics.

\subsection{Mean-field theory links the emergent threshold behavior to a phase transition}

\begin{figure*}[htp]
    \centering
    \includegraphics[width=\textwidth]{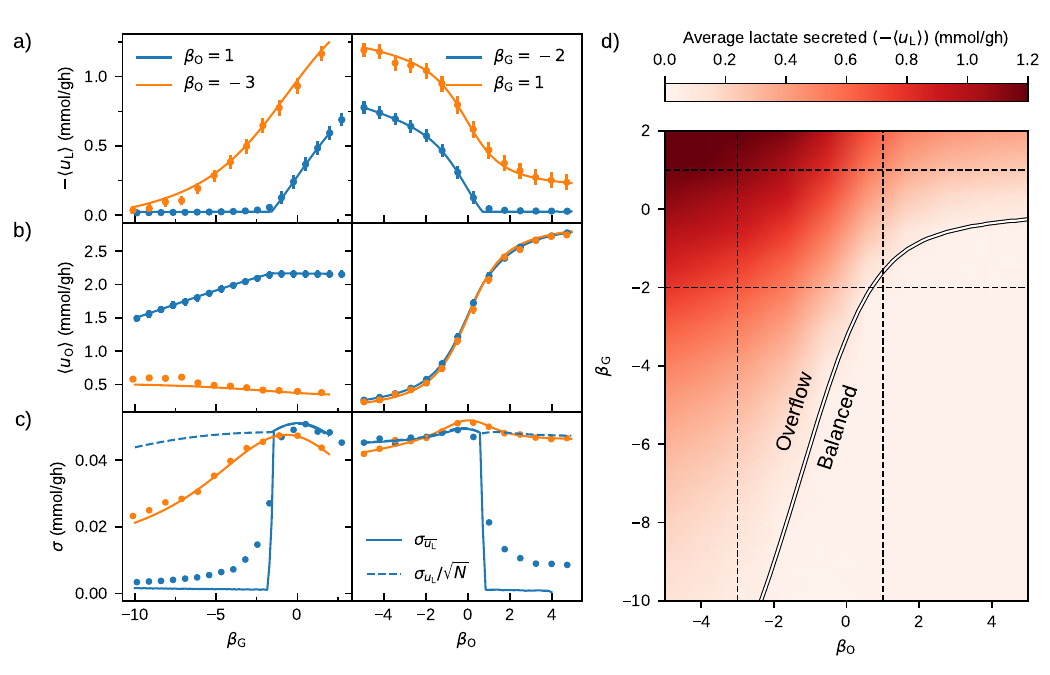}
    \caption{
    Mean-field approximation of the CBM simulations.
    Comparison between mean-field analytics (lines) and numerical simulations (points) for (a) the mean lactate flux ($-\expect{u_\lact}$), (b) mean oxygen flux ($\expect{u_\ox}$), and (c) standard deviation of the mean lactate flux ($\sigma_{\bar{u}_\lact}$\added{, the dashed lines stand for the single cell standard deviation}) as functions of $\beta_\gluc$ at fixed $\beta_\ox$ (left), and of $\beta_\ox$ at fixed $\beta_\gluc$ (right). Error bars everywhere indicate the standard error on the quantities. Simulations were performed by sampling the feasible space $\mathcal{F}_N$ of $N=150$ cells spread in an area of $500 \times \qty{500}{\micro\meter\squared}$ according to (\ref{eq:boltz2}) via Hit-and-Run Monte Carlo Markov chains. Analytical lines were obtained by solving the mean-field model (\supmethod{3}).
    (d) Mean-field phase diagram in the plane spanned by $\beta_\ox$ and $\beta_\gluc$. The white line is a line of phase transitions where average flux variances are discontinuous (see panel (c)), and separates the `overflow phase' (above the line) from the `balanced phase' (below), the two differing by the rate of lactate accumulation in the medium (background color scale).
    }
    \label{fig:meanfield}
\end{figure*}

To \deleted{further} elucidate the behavior uncovered by sampling  $\mathcal{F}_N$ for varying $\beta_\gluc$ and $\beta_\ox$, we analyzed a mathematically solvable approximation of our constraint-based model. To define it, we focused in particular on the set of constraints (\ref{eq:diffusion}), which distinguishes $\mathcal{F}_N$ from $N$ independent copies of the single-cell space $\mathcal{F}_1$. Upon isolating the contribution of cell $i$ ($i=1,\ldots,N$), the diffusion constraints (\ref{eq:diffusion}) can be re-cast as
\begin{equation}
u_\lact^{(i)} + \origsum_{j\neq i} A_{ij} u_\lact^{(j)}\leq 0\qquad (i = 1,2,\ldots,N)~~.
\end{equation}
The sum on the left-hand-side depends on the specific values of the fluxes $u_\lact^{(j)}$ as well as on the relative positions of cells (see (\ref{eq:Aij})).
\replaced{For the sake of tractability, we however assume $N\gg 1$ and replace all coefficients $A_{ij}$ with an $N$-dependent constant factor $K/N$, thereby discarding effects due to the spatial organization of cells so that all pairs of cells interact with the same strength.
To estimate $K$, we note that, upon neglecting fluctuations,}{In particular, we set $A_{ij}=K/N$ for all $i\neq j$, where $K$ is a factor that can be easily estimated for $N\gg 1$ upon neglecting fluctuations. Indeed from}
\begin{equation}
    \origsum_{j\neq i}A_{ij}u_\lact^{(j)}\simeq K \avgul ~~,
\end{equation}
where $\avgul$ denotes the  mean lactate flux across cells (\added{coinciding with the ensemble average $\langle u_\lact \rangle$ for large $N$, see also \supmethod{3})}, and
\begin{equation}
    \origsum_{j\neq i}A_{ij}u_\lact^{(j)}
        \simeq
        N \avgul \frac{R}{d}
        =
        \alpha RL \rho \avgul~~,
\end{equation}
with $d$ the mean cell-to-cell distance in a square of size $L$, $\rho=N/L^2$ the density of cells, and $\alpha\simeq 1.918$ a numerical constant. \replaced{It then follows that}{, one finds} $K\simeq \alpha RL\rho$ (see \supnote{1} for the actual numerical values).
Within this approximation, \replaced{we therefore re-write (\ref{eq:diffusion}) as}{we can replace  with}
\begin{equation}
\label{eq:diffusion_mean}
    u_\lact^{(i)} + K\avgul \leq 0\qquad (i = 1,2,\ldots,N)~~.
\end{equation}
\replaced{One sees that, from a physical viewpoint, our choice is}{Physically, this is} equivalent to assuming that the lactate flux of every cell in the population is coupled to a bulk lactate flux to which all cells contribute. \replaced{This is known in physics as a `mean-field approximation'}{ known as a `mean-field approximation' in statistical physics} (see e.g. \cite{parisi1998statistical}, Ch.\,3).

A detailed examination of the mean-field \replaced{model}{approximation from this point of view} is presented in \supmethod{3}, including the analytical solution in the limit $N\to\infty$. Crucially, in this approximation the constraints (\ref{eq:diffusion}) become formally identical for all $i$'s (see (\ref{eq:diffusion_mean})), which effectively reduces the study of the multi-cellular space $\mathcal{F}_N$ to that of the single-cell space $\mathcal{F}_1$ supplemented with the additional constraint (\ref{eq:diffusion_mean}).
The mean field approximation leads in turn to \deleted{two} non-linear \replaced{self-consistency}{self-consistent} equations for two emerging order parameters, one of which can be identified, in the thermodynamic limit, with the average net lactate flux $\langle u_\lact \rangle$. \added{The equations admit an explicit analytical solution only for $\langle u_\lact \rangle<-U_\ox/(3K)$; for larger $\expect{u_\lact}$ they have to be solved numerically. The presence of background lactate can be accounted for by adding a constant to the term $K \avgul$ in the equations (\supmethod{4}).}

\replaced{The numerical solution for $\langle u_\lact\rangle$ versus $\beta_\gluc$ at fixed $\beta_\ox$}{Our analytical expression for $\langle u_\lact \rangle$ is found to define an ``overflow curve'' that quantitatively reproduces the sampling results of the previous section (see Fig.~\ref{fig:meanfield}A, compare with Fig.~\ref{fig:sampling}C).} quantitatively reproduces the sampling results of the previous section (see Fig.~\ref{fig:meanfield}a, compare with Fig.~\ref{fig:sampling}c).
\replaced{Furthermore, we found that}{It turns out that the equations admit an explicit analytical solution as soon as $\langle u_\lact \rangle<-U_\ox/(3K)$ while beyond the threshold have to be solved numerically  (see \supmethod{3}). More precisely, in the regime where the average lactate flux of the population displays threshold-linear behavior,} the derivative of $\expect{u_\lact}$ displays a discontinuity at the onset point of overflow metabolism. \replaced{The values of $\beta_\ox$ and $\beta_\gluc$ where the discontinuity occurs define}{As displayed in Fig.~\ref{fig:meanfield}B. (top panel), this in turn defines (\supmethod{3})} a curve in the $(\beta_\ox,\beta_\gluc)$ plane that separates a phase with large lactate spillover (`overflow phase', above the curve) from one without (`balanced phase', below the curve).
\added{Such a curve is displayed in Fig.~\ref{fig:meanfield}d.}
Points along the curve denote critical values of $(\beta_\gluc, \beta_\ox)$ corresponding to the transition between the two regimes.
The comparison between numerical simulations and the mean-field analytical predictions shows an excellent quantitative agreement for the flux averages over the whole range of parameters, as well as for the average flux fluctuations ($\sigma_{\bar{u}_\lact}$) in the unbalanced phase.
However, mean-field theory underestimates fluctuations below the phase transition, \deleted{where the agreement is only qualitative,} see Fig.~\ref{fig:meanfield}c. 
This is an expected \deleted{common} pitfall of mean-field approximations, \replaced{and more refined}{across many system, and a better characterization of phase transitions in finite dimension  would require devoted} combinatorial \cite{baxter2016exactly} or field-theoretic \cite{parisi1998statistical} calculations \added{would be required to overcome it}.

\replaced{A key}{One important} difference between the two phases is the presence of negative inter-cellular flux correlations in the balanced phase that are absent in the overflow phase, where  fluctuations approximately follow the law of large numbers ($\sigma_{\overline{u}_\lact}\simeq \sigma_{u_\lact}/\sqrt{N}$,
\added{as it can be seen from continuous vs dashed blue lines in Fig.~\ref{fig:meanfield}c, see also \supmethod{6}).}
Our \replaced{results therefore support}{analytical results therefore provide further quantitative support to} the idea that the management of lactate levels in the medium is a genuine emergent phenomenon, achieved through a population-level coordination of lactate exchange fluxes. Likewise, coordination failures triggered by small changes in $\beta_\gluc$ and/or $\beta_\ox$ can lead to excess accumulation of extra-cellular lactate. \deleted{a characteristic shared with other systems that undergo order-disorder phase transitions.}

\subsection{Inverse modeling experimental data%
}

\begin{figure*}[htp]
    \centering
    \includegraphics*[width=\textwidth]{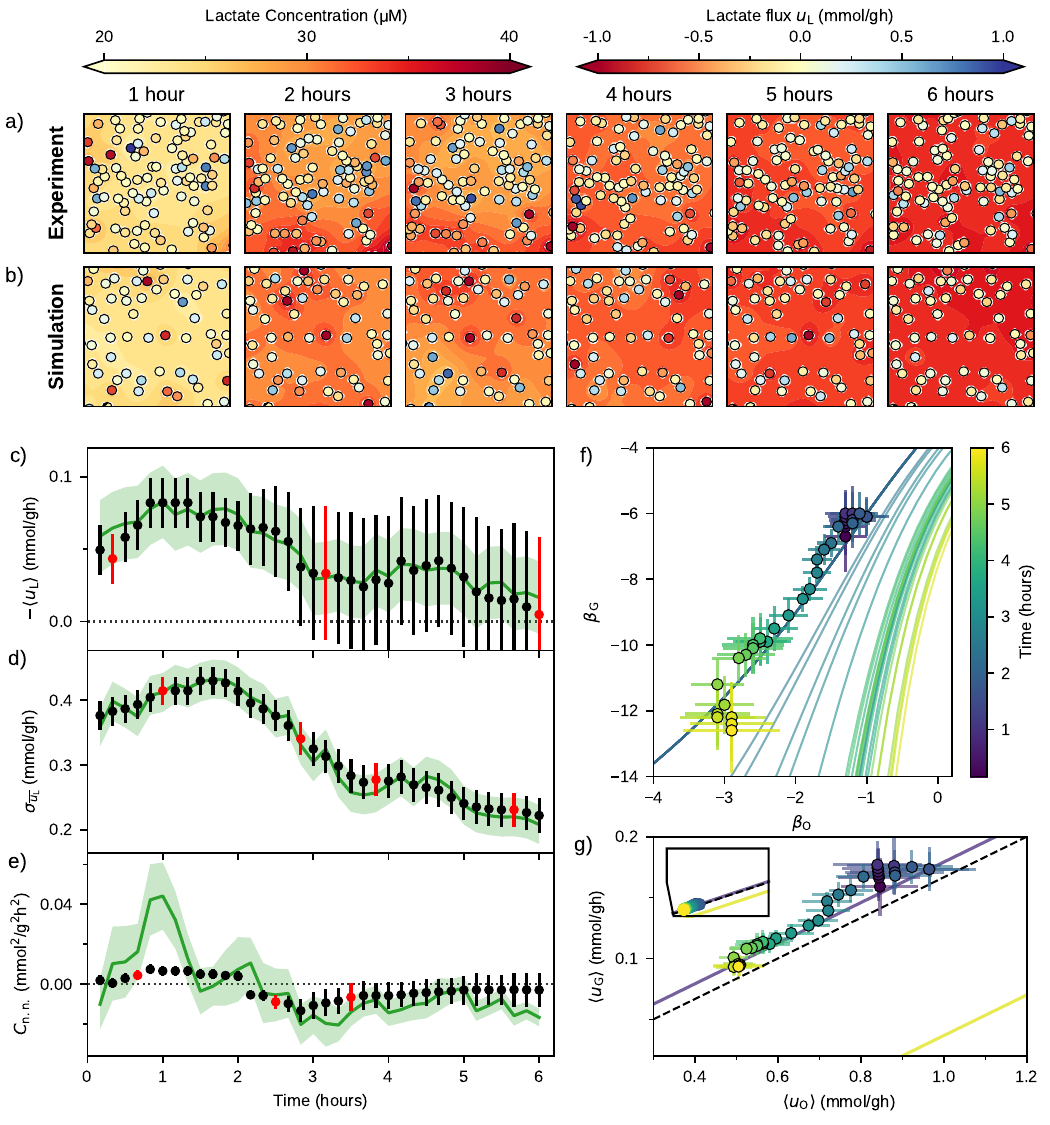}
    \caption{Comparison between theoretical results and empirical data.
    Snapshots of lactate gradient and single-cell fluxes from (a) experimental frames of Ref. \cite{onesto2022probing} (at intervals of 1 hour) and (b) from CBM simulations performed by sampling (\ref{eq:boltz2}) with parameters $\beta_\gluc$ and $\beta_\ox$ that, at each time step, provide the maximum likelihood reconstruction of the empirical average lactate flux. See \supmethod{8} for details.
    \deleted{Comparison between empirical (bars) and theoretical (lines) single-cell lactate flux distributions at different times for coarse-grained frames.}
    \added{Comparison between empirical (line: mean value, shaded region: standard error from jackknife resampling) and theoretical (error bars; black: fit, red: predictions) flux values as a function of time for (c) mean lactate flux, (d) standard deviation of single cell lactate flux and (e) nearest-neighbor correlations. The errors on the inferred quantities were calculated from the errors on the inferred values of $\beta_\gluc$ and $\beta_\ox$.}
    (f) Mean-field phase diagram 
    with the critical line\replaced{s and the}{ in white, together with} inferred values of $\beta_\ox$ and $\beta_\gluc$ (markers with error bars) colored according to the time stamp for the 36 frames of the experimental dataset. For details on how the errors were estimated, see \supmethod{8}. \deleted{Estimated contour levels of average pH are provided.}
    (g) Same as (f), but in the plane ($\expect{u_\ox},\expect{u_\gluc}$). \replaced{Only the first and last critical lines are shown. The}{The critical line is again drawn in white, and the} dashed black line corresponds to $\expect{u_\lact}=0$.
    (Note that the latter line does not appear in (f).)
    Inset: zoomed out view in the single-cell flux space $\mathcal{F}_1$ of Fig.~\ref{fig:network}b. }
    \label{fig:inference}
\end{figure*}

To compare our theoretical scenario with actual experimental data, we focused on recent experiments characterising the dynamical pH landscape in co-cultures of human pancreatic cancer cells (AsPC-1) and cancer-associated fibroblasts \cite{onesto2022probing}. Advances in nanofibers technology have now made it possible to probe the cellular microenvironment in cultures at high spatial and temporal resolution \cite{moldero2020probing}. In the dataset we considered, local pH data were collected every 10 minutes  over a 6-hour timespan, yielding 36 snapshots of the population's adaptation (lag phase) to the culture medium.
Time-resolved \added{estimates for} single-cell lactate fluxes \replaced{have been previously calculated from  extracellular proton levels}{for these cultures have been previously obtained by inverse-modeling extracellular proton levels at each time frame} (see \cite{onesto2022probing} and \supmethod{7}). A highly heterogeneous \added{flux} profile \deleted{of lactate fluxes} was found, \deleted{which in turn underlies} \added{suggesting} a complex \added{underlying} lactate exchange network.

We used these \replaced{single-cell flux data to inform a constrained maximum-likelihood inference problem returning the values of $\beta_\gluc$ and $\beta_\ox$ that, at each time frame, yield the best prediction of the  empirical bulk lactate flux, the single-cell standard deviation, and correlations between the lactate fluxes of nearest-neighbor cells, i.e.
\begin{equation}
    C_{\mathrm{n.n.}} = \frac{1}{N} \origsum_{i=1}^N u_\lact^{(i)} u_\lact^{(n(i))} ~~,
\end{equation}
where $n(i)$ denotes the nearest neighbor of cell $i$. The inference pipeline is described in detail in \supmethod{8}. $C_{\mathrm{n.n.}}$ plays an especially important role. In a system of non-interacting cells behaving independently, $C_{\mathrm{n.n.}}$ would be constrained to be positive definite (and equal to the square of the average flux, see \supmethod{6}).
Empirical values of $C_{\mathrm{n.n.}}$  however strongly deviate from this expectation both in absolute value and in sign (\supmethod{6}), thereby stressing the need for an interacting model.
Further, a non-interacting model made of isolated single cells is unable to fit experimental fluctuations and averages of the lactate fluxes given its strongly constrained scaling behavior (\supmethod{8}).
Global diffusion constraints like (\ref{eq:diffusion}) indeed enable for  both negative correlations and larger absolute values thereof (\supmethod{8}). To obtain quantitative agreement, however, we found that one must account for two additional ingredients.
First is the lactate accumulating in the medium over time, which, as said above, alters the form of (\ref{eq:diffusion}) (\supmethod{4}). The net effect of a constant background term in the mean-field model is a downward shift of the line separaring the Warburg from the balanced regime in the $(\beta_\ox,\beta_\gluc)$ plane (\supmethod{4}).
In experiments, the bulk lactate level was found to increase over time at a rate that decreases in time  \cite{onesto2022probing} (See also \supmethod{7}). In such a scenario, the critical line of the mean-field model is expected to move downward in time. During most of the experiment, the empirical bulk lactate level appears to have a small impact on the population, as e.g. $C_{\mathrm{n.n.}}$ stays positive.
In the last two hours, however, when the background lactate concentration exceeds around $\qty{30}{\micro M}$, $C_{\mathrm{n.n.}}$ becomes negative (implying net exchanges between neighbors) with a small net lactate intake. The latter is at odds with empirical data, which display  that lactate is on average still excreted, albeit at reduced rates (\supmethod{7}).
We therefore solved this quantitative inconsistency by adding to the likelihood function a small phenomenological constant $J>0$ coupling nearest neighbors, which amounts to including a factor of the form
\begin{equation}
    \exp \left[ -J \origsum_{i=1}^N u_{\lact}^{(i)} u_{\lact}^{(n(i))} \right]
\end{equation}
in the Boltzmann weight. While it is necessary to fully recover empirical data, this term only causes a small perturbation when accounted for in the mean-field model (\supmethod{6}).}{These are the only two fitted parameters. We then sampled the distributions with the inferred values of $\beta_\gluc$ and $\beta_\ox$ again using Hit-and-Run Monte Carlo, and compared the results against the experimental time course. }

\added{To summarize, the inference protocol is based on three fitting parameters, namely $\beta_\ox$ and $\beta_\gluc$ (which take a different value for each time frame), plus the coupling constant $J>0$ (which instead is the same for all frames). The complete likelihood function is reported in \supmethod{10},  along with the details of the inference pipeline.}

\replaced{Results are reported in Fig.~\ref{fig:inference}. Panels (a) and (b)}{Fig.~\ref{fig:inference}A} display a comparison between lactate levels and fluxes from empirical data and simulations respectively \added{(obtained by sampling the inferred model)} over time, showing for simplicity a reduced set of 6 frames \replaced{(every hour).}{derived by coarse-graining the empirical dataset of 36 snapshots over time by averaging over 6 snapshots (corresponding to about 1 hour) per coarse-grained frame. }
Besides the qualitative frame-by-frame agreement, one \replaced{can see}{sees} how inferred models \added{quantitatively} capture the dynamics of the culture
\replaced{by comparing empirical and model-derived time trends of bulk lactate flux (Fig.~\ref{fig:inference}c), single cell variability (Fig.~\ref{fig:inference}d) and nearest-neighbor correlations (Fig.~\ref{fig:inference}e). Note that, in order to appraise the predictive capabilities of the model, data were divided into a training set and a test set (ratio 9 to 1). A very good agreement is found, for both the training and test sets (reduced $\chi_{\mathrm{training}}^2 \simeq \chi_{\mathrm{test}}^2 \simeq 0.8$).}{from an initial state characterized by high environmental lactate levels and strongly heterogeneous single-cell fluxes to a final state with low environmental lactate and reduced flux variability.  This point becomes quantitatively clear by comparing empirical distributions of lactate fluxes to theoretical ones (Fig.~\ref{fig:inference}B).}

We next mapped the values of $(\beta_\ox,\beta_\gluc)$ that provide the best fits over time onto the phase diagram obtained from the mean-field model (Fig.~\ref{fig:inference}\replaced{f}{B top}).
\deleted{Remarkably, inferred points consistently map around the critical line associated with high inter-cellular exchange and separating the overflow phase from the balanced phase (Fig.~4C).}
Following a transient, the points display a clear time-ordering \replaced{(i.e. a rather well defined trajectory in the $(\beta_\ox,\beta_\gluc)$ plane), which}{that} suggests a strong population-level regulation of both $\beta_\gluc$ and $\beta_\ox$ toward negative values\added{: over time, the population appears to reduce its overall metabolic rate}.
A different representation of the same result is given in Fig.~\ref{fig:inference}g, where the average net glucose and oxygen fluxes ($\expect{u_\gluc}$ and $\expect{u_\ox}$, respectively) are used as control parameters instead of $\beta_\gluc$ and $\beta_\ox$.
\deleted{Here one sees that inferred points lie in a region of phase space without deleterious lactate accumulation in the extra-cellular medium with ensuing acidification as it can be seen from the contour levels of the estimated pH in Figs. 4C and D (estimated from the lactate level Sec.S7).}
\replaced{In short, as time progresses and lactate accumulates in the medium, the population exhibits greater metabolic coordination (reflected in negative correlations), so that the spillover reduces in time}{In short, there is no significant lactate accumulation in the culture} in spite of the presence of \deleted{single} cells that sustain large lactate export fluxes.
\replaced{In other terms, carbon overflow}{(i.e. CO} at the level of single cells does not necessarily imply environmental spillover of lactate.
\replaced{Furthermore,}{) Taken together, these results imply that the degree of inter-cellular lactate exchange is coordinated over time to achieve a configuration that lies close to the critical threshold of the initial point (roughly at zero overflow), but fail to reach the critical point of the current time, that would imply a net intake. More specifically,} since both the inferred $\beta_\gluc$ and $\beta_\ox$ tend to decrease in time as adaptation to the medium progresses, this population appears to move in the single-cell space $\mathcal{F}_1$ close to the $u_\lact=0$ line and (roughly) toward the state of optimal ATP yield \deleted{in the balanced regime} (see Fig.~\ref{fig:inference}g versus Fig.~\ref{fig:network}b).
\added{However, because the threshold for the balanced phase shifts over time due to accumulating lactate, the population, while reducing lactate spillover, remains in a Warburg regime throughout the dynamics.}

\subsection{Mitochondrial saturation versus local hypoxia: dynamics of oxygen usage}

\begin{figure*}[htp]
    \centering
    \includegraphics[width=\textwidth]{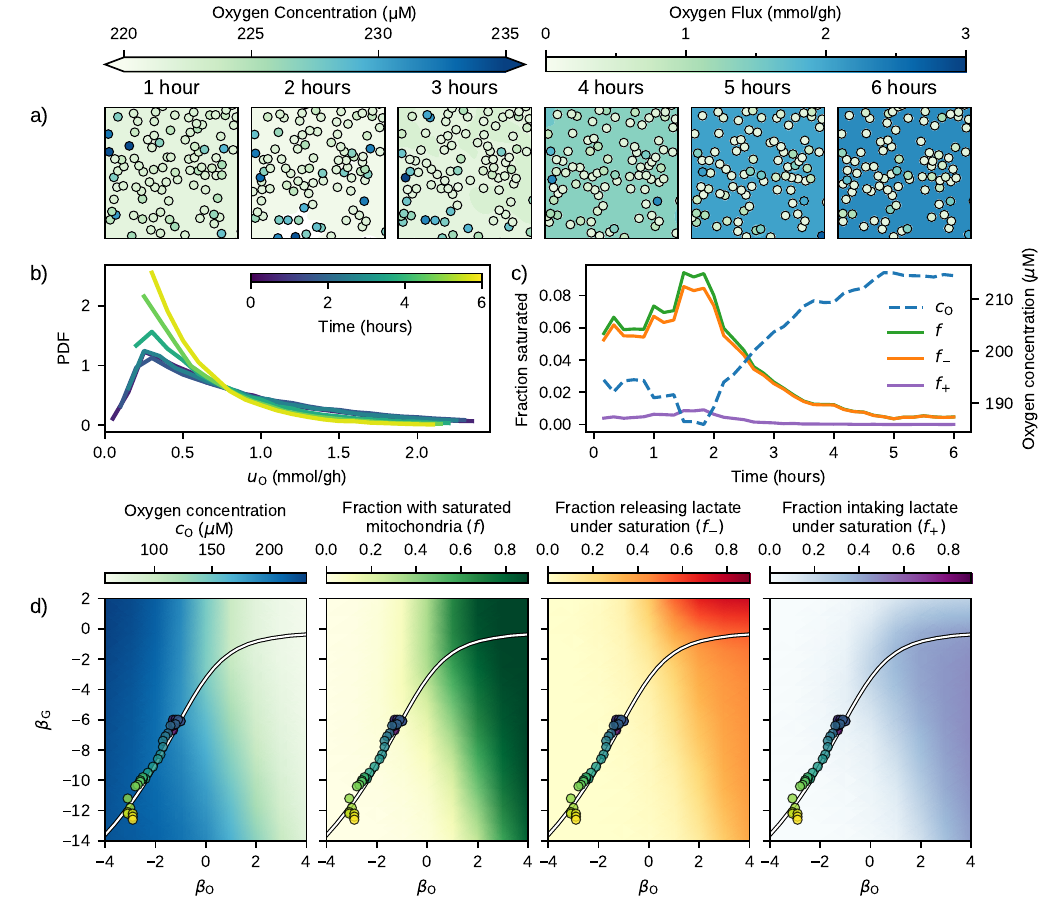}
    \caption{Inferred dynamics of oxygen usage.
    (a) Snapshots of oxygen levels and single-cell fluxes inferred from \deleted{coarse grained} experimental frames (at intervals of 1 hour). Note that oxygen fluxes cannot exceed stamps the value $U_\ox = \qty{3}{\mmolgh}$ (see (\ref{eq:u_ox})). See \supnote{1} for details.
    (b) Normalized histogram of inferred single-cell oxygen fluxes at different times from coarse-grained snapshots.
    (c) Inferred time course for bulk oxygen level ($c_\ox$), fraction of cells with saturated oxidative capacity ($f$), and fraction of saturated cells with lactate release ($f_-$) and import ($f_+$). Note: $f=f_-+f_+$.
    (d) Mean-field phase diagram 
    in the $(\beta_\ox,\beta_\gluc)$ plane plotted using the corresponding values of $c_\ox$, $f$, $f_-$ and $f_+$ as a background heat map. The critical line is in white, while values of $\beta_\ox$ and $\beta_\gluc$ (markers) inferred from the 36 frames of the experimental dataset of \cite{onesto2022probing}  are colored according to their time stamp. For visual clarity, only the first critical line is shown.}
    \label{fig:ox}
\end{figure*}

In light of the above results, it would be important to understand whether, in the observed scenario, lactate \replaced{excretion}{production} by individual cells is triggered by saturated mitochondrial capacity (which diverts excess carbon towards fermentation) or, rather, by local hypoxic conditions (which limit the oxidative processing of nutrients).
\added{Within the framework of our model,  cell $i$ has saturated mitochondrial capacity if its oxygen import flux $u_\ox^{(i)}$ is close to the limit $U_\ox = \qty{3}{\mmolgh}$ (see (\ref{eq:u_ox})); on the other hand, it suffers from local hypoxic conditions if it experiences a local concentration of oxygen so small that its oxygen import flux becomes diffusion-limited.
In this case, a diffusion constraint for oxygen would kick in, of the form
\begin{equation}
    u_\ox^{(i)} + \origsum_{j\neq i}\frac{u_\ox^{(j)}}{|\mathbf{r}_i-\mathbf{r}_j|/R}
    \simeq
    4\pi D_\ox R \,c_{\ox,\infty} ~~,
\label{kgjfhd}
\end{equation}
with $c_{\ox,\infty}\simeq \qty{250}{\micro M}$ the background oxygen concentration (Eq. (\ref{kgjfhd}) expresses that fact that the concentration of oxygen at the boundary of cell $i$ is $c_{\ox}^{(i)}\simeq 0$, see \supnote{2}).}

In absence of high-resolution data about local oxygen levels in the culture,  addressing this question requires an inference framework that goes from learning distributions to learning single-cell fluxes. \deleted{To this aim,} We can \added{however} use single-cell lactate fluxes derived from the dataset of \cite{onesto2022probing}, along with the estimate for the average oxygen flux derived above, to obtain a prediction for single-cell oxidative fluxes via a Boltzmann sampling of the $N$-cell space of feasible flux configurations $\mathcal{F}_N$.
Theoretical and computational details of the method employed are given in \supmethod{9}.  While \deleted{we are not able to infer} single cell oxygen fluxes \replaced{show large uncertainties due to the ``sloppiness''\cite{gutenkunst2007universally} of the inferred model}{unambiguously}, we can reconstruct a plausible scenario whose key results are given in Fig.~\ref{fig:ox}.

We first notice (Fig.~\ref{fig:ox}a) that, due to a faster diffusion rate, oxygen profiles across the culture appear much more homogeneous than lactate profiles.
In addition, single-cell fluxes display a time trend towards downregulation and reduced heterogeneity, a pattern consistent with the time-course of $\beta_\ox$ shown in Fig.~\ref{fig:inference}f.
This is quantified by how the normalized histogram of single-cell $u_\ox$ values shifts over time (Fig.~\ref{fig:ox}b). Cells sustaining an oxygen import flux closer to the saturation point ($U_\ox \sim \qtyrange{2}{3}{\mmolgh}$ , see (\ref{eq:u_ox})) become more and more rare as time progresses.

A closer look at the time course (Fig.~\ref{fig:ox}c) highlights two distinct regimes in the population's dynamics. \added{While remaining consistently below 10\%}, the fraction $f$ of saturated mitochondria initially increases as cells seem to increase oxygen consumption (consistently with the transient increase of $\beta_\ox$ that is visible in Fig.~\ref{fig:inference}f), leading to a decrease of average environmental oxygen levels $c_\ox$.
After about 2 hours, $f$ inverts the trend and begins to decrease while $c_\ox$ concomitantly increases, signaling that cells have stabilized their metabolism at reduced levels of lactate export, import and exchange (Figs. \ref{fig:inference}a--d).
Note that mitochondrial saturation can occur both under excess glucose intake, leading to the release of lactate in the medium, and under intake of lactate from the medium. We denote by $f_-$ (resp. $f_+$) the fraction of cells in the former (resp. latter) condition. Fig.~\ref{fig:ox}c shows that, while (perhaps surprisingly) saturation with lactate import is more common, both $f_+$ and $f_-$ follow the same time trend as $f$.

Overall, these results, along with evidence that oxygen concentration consistently stays above typical hypoxia levels ($\qtyrange[per-mode=symbol,range-units=single]{2}{3}{\mg\per\liter} \text{ of \ce{O2}} \simeq \qty[per-mode=symbol]{60}{\micro\mole\per\liter}$ \cite{herman2016physics}), \added{imply  that the inferred single cell oxygen fluxes stay well above the diffusion limit in our experimental conditions (see also \supmethod{5}). This} suggests that the observed dynamics is more likely coupled to the collective behavior of the population and the emergent exchange network than to a depletion of local environmental oxygen\deleted{ availability}.

To further support this \replaced{picture}{idea}, it is \replaced{useful}{instructive} to revisit the phase structure of Fig.~\ref{fig:inference}f \replaced{against}{on} the background of the quantities used to analyze oxygen dynamics, reported in detail in Fig.~\ref{fig:ox}d.
While macroscopic features appear to be strongly tied to the value of $\beta_\ox$, one notices that the region where acidification is most severe ($\beta_\gluc>0,\beta_\ox<0$) is not generically associated to hypoxic conditions or mitochondrial saturation and, again, both conditions characterize the region where acidification is mild ($\beta_\gluc<0,\beta_\ox>0$). This shows that single-cell and population-level features are separated, the latter being essentially driven by inter-cellular interactions.

\section{Discussion}

The broad biological question we faced here asks whether the metabolic phenotype of a multi-cellular system emerges from the interactions among its individual components or is rather the result of a multitude of independent cell-autonomous behaviors.
The results we present support the idea that metabolic interactions between cells play a central role in shaping their shared environment, thus influencing the overall fitness of the population.
More precisely, cells appear to collectively regulate the levels of medium-acidifying compounds through time-dependent coordination of their exchanges with the surroundings. This process is ultimately sustained by the establishment of cell-to-cell interactions facilitated by the transport of overflow products.
In the presence of coordination, environmental spillover is limited despite the presence of cells exporting these compounds at high rates.
Conversely, the accumulation of compounds in the medium can be seen as a \replaced{failure to coordinate}{breakdown of  coordination}.

Our findings \added{furthermore} provide robust evidence that the large-scale metabolic organization of cell populations exhibits hallmarks of phase transitions\replaced{, and may therefore}{. Therefore, these phenomena can} be understood, and possibly controlled, through the application of concepts derived from statistical physics.
\added{Indeed,} when viewed through the lens of \replaced{collective phenomena}{statistical physics}, the \replaced{crossover}{transition} from a balanced to an unbalanced state with overflow metabolism bears significant similarities to
\replaced{the standard disorder-to-order transitions that occur, for example, in magnetic systems. These models have}{indeed} shown that highly non-trivial macroscopic properties
\deleted{(like phase transitions, symmetry breaking, universality, etc.)} can arise from the interaction patterns of large assemblies of simple, identical
\replaced{variables}{magnetic spins} \cite{nishimori2011elements}.
\replaced{In recent years, many of the methods and insights developed for the study of these systems have been ported to other disciplines, including ecology  \cite{advani2018statistical, altieri2019constraint, batista2021path}. In view of its focus on inter-cellular couplings and their consequences, our work is indeed close to ecological settings, albeit perhaps}{In the present case, by using the statistical physics toolbox, we have derived a theoretical `critical line' that separates regimes in the plane spanned by control parameters directly related to the average glucose and oxygen intakes by the population. Purely theoretical studies of phase transitions and applications of statistical physics methods to ecological settings have been recently put forward and our work support this approach providing quantitative empirical evidence} on the more basic ground of cellular metabolic dynamics.

\replaced{Finally, our results}{On the theoretical side, our research thus} confirm that heterogeneous cell populations can be \deleted{well} described \added{effectively} by maximum-entropy distributions like (\ref{eq:boltz}) \cite{de2016growth, de2018statistical, fernandez-de-cossio-diazCharacterizingSteadyStates2017, cossioS,cossioJij, tourigny2020dynamic, rivas2020metabolic}, suggesting that, at least in some conditions, the constrained maximisation of population-level variability might be a reasonable objective for multi-cellular systems.
\deleted{Moreover, the framework developed here requires inferring only a small number of parameters and constraints (two per time frame) to effectively describe the metabolism of large, spatially organized populations over time.
This stands in contrast to the complex task of inverse modeling spiking data using Ising neural networks, which demands inferring thousands of parameters, one for each synap tic connection.
Furthermore, an inverse-modeling scheme has shown that experimental data obtained from real cell populations seem to remain close to this critical line as they adapt over time to a given environment.
The observation that cultures appear to adapt to a medium by remaining close to a critical line could have important biological implications, since even small changes in the mean import flux of glucose or oxygen, could have drastic effects on cell-to-cell variability and, in turn, on the ability of the population to regulate its environment. }

\deleted{This leaves an important open question, namely why would cell populations tend to remain close to critical in these conditions? In brief, our findings indicate that, by remaining near criticality, a population can curb the environmental costs of carbon overflow while maintaining a large (albeit not maximal) degree of cell-to-cell heterogeneity, for instance in terms of glucose import fluxes, oxygen import fluxes and lactate exchanges.
In addition, such a balance appears to be implemented over time, as the population improves the energy yield. The benefits of reducing environmental costs, i.e. pollution of the medium, are self-evident. It is perhaps less clear why a more heterogeneous state (i.e. the critical state) should be preferred to a more homogeneous one (in the balanced phase).
A possible explanation lies in the regulatory costs associated to cell-to-cell variability. Among equally fit populations, those with reduced levels of heterogeneity are necessarily subject to tighter regulatory constraints compared to more variable ones, and therefore face larger costs to implement these constraints population-wide.
It is therefore reasonable to think that populations that evolve in time to maximize some context-dependent objective function (e.g. growth rate, energy yield, etc.) do so while keeping these regulatory costs as small as possible.
In the present case, the performance of the population is affected by two types of costs: one related to lactate spillover, the other due to reduced variability. As the population evolves in time towards an energy-yield optimum, and both $\beta_\gluc$ and $\beta_\ox$ decrease, so does cell-to-cell variability.
Critical or near-critical states however appear to consistently strike the optimal balance: above the critical line, environmental costs increase too much; below it, regulatory costs increase too much. All in all, these observations are in line with the idea that `critical states' arising from interactions among their constituents are often crucial to understand the emergence of non-trivial large-scale behavior in living systems 
Going beyond our case study, implementing this trade-off as a type of programmed cellular behavior could lead to efficient division of labor within tissues that would confer a selective advantage to multi-cellular organisms.}

\replaced{On the more biological side, the inverse-modeling scheme we developed indicates quantitative ways to appraise the adaptive metabolic response of populations of cells to their environment, with individual-cell resolution.
As a plus, our protocol requires inferring only a small number of parameters and constraints in order to effectively describe the metabolism of large, spatially organized populations over time.
This stands in contrast to the complex task of inverse modeling spiking data using e.g. Ising neural networks, which typically demands the inference of one parameter per synaptic connection \cite{schneidman2006weak}.
Upon analyzing experimental data from real tumor-stroma cell cultures during the adaptation phase, we found that cells coordinate their metabolism and reduce their metabolic rate over time to prevent environmental deterioration caused by lactic acid overflow.
Further analysis indicates that individual cells operate far from mitochondrial saturation and well below the oxygen diffusion limit (i.e. away from local hypoxia). Overall, this picture strongly suggests that the primary driver of the observed population-level dynamics is not to be found in metabolic constraints but, rather, in a failure of cellular coordination.
We showed that a version of the model made of isolated single cells (the usual assumption of flux balance analysis  models \cite{orth2010flux}) is unable by construction to retrieve observed negative cell-cell correlations and the fluctuations scaling  of metabolic fluxes (\supmethod{8}).
Unfortunately, such a scenario could only be}{This scenario admittedly has only been} validated using a single dataset, and should \added{ideally} be contrasted with \deleted{new evidence coming from} single-cell flux data \replaced{taken in}{across} different experimental conditions (e.g. different cell populations, density, geometry and medium composition).
Although recent advancements in high resolution mass-spectrometry hold promise \cite{wang2022spatially, rappez2021spacem}, they are currently expensive and invasive.
On the other hand, the very recent development of micro-environmental sensing \replaced{protocols}{methods} \cite{rizzo2023ph, grasso2023fluorescent, feng2024severely, grasso2024highly}, coupled with ad hoc inverse modeling schemes, has the potential to lead to fully time-resolved single-cell metabolic flux analysis.

\replaced{Besides the highly simplifying assumptions on which it relies, the major limitation of our theoretical work lies, we believe, in its static (quasi-equilibrium) character.
A constant background lactate level allows for the analytical tractability that, in turn, unravels the connection between overflow phenomena and phase transitions. Empirical data are however not stationary.
In our case study, this aspect appears to become especially important only in the final part of the adaptation dynamics. To make}{In order to extend} our \replaced{theoretical}{statistical physics} framework \added{applicable} to more general \deleted{experimental} data\added{sets}, it will \added{however} be important to \deleted{go beyond the quasi-equilibrium assumption implicit in our maximum entropy scheme, allowing} explicitly \added{consider the} coupled dynamics \replaced{of}{between} the state of environment and the behavior of cells.
This could \replaced{generalize models of population}{extend current efforts on} metabolism that \replaced{rely on the}{makes the simplifying} assumption of single cell optimality (like COMETS  \cite{dukovski2021metabolic}).
\replaced{Such an}{and whose} implementation could benefit from recent generalization of the maximum-entropy scheme in out-of-equilibrium settings known as maximum-caliber \cite{ghosh2020maximum}.  \deleted{A  dynamical theory would also help to tell apart  self-organized criticality from the one observed in coarsening dynamics}.

\replaced{In addition to}{Besides} fundamental aspects, our \replaced{work}{scenario} offers \deleted{immediate} insights in the context of tumor metabolism. In higher organisms, where metabolic behavior is specialized and compartmentalized according to cell type, the exchange of metabolites such as lactate occurs across multiple scales: organs, tissue and single cells \cite{brooks2009cell}.
This division of labor \replaced{through}{by} metabolic specialization and exchange provides significant benefit to healthy tissues. However, it also renders these systems vulnerable to exploitation by malignant cells, which undergo metabolic rewiring during carcinogenesis \cite{pavlova2022hallmarks}.
As first observed by Warburg in the 1920s \cite{Warburg1925}, tumours exhibit a notable tendency to consume excessive amounts of glucose while producing lactate, even in the presence of oxygen.
However, despite extensive research efforts, a comprehensive understanding of the Warburg effect in cancer has remained elusive to date \cite{deberardinis2020we}.
The approach presented herein supports the idea that the Warburg effect may reflect an emergent feature of a large population of interacting cells characterized by a highly heterogeneous pattern of lactate exchange among individual cells.
This in turn would place considerable weight on the ecological dynamics of tumor development, particularly in its early stages. Such a scenario would be fully consistent with recent experimental findings \cite{hensley2016metabolic}.

\section{Methods}

\subsection{Model and simulation parameters}

The model parameters were chosen to closely match the experimental settings of \cite{onesto2022probing}, with approximately 40000 cells of radius \qty{10}{\micro\meter} and dry weight \qty{1}{\nano\gram} randomly distributed in an area of \qty{1}{\centi\meter\squared}, with background glucose ($c_\gluc$) and oxygen ($c_\ox$) concentrations \qty{25}{\milli\molar} and \qty{0.25}{\milli\molar} respectively. The simulations were performed in a window of \qtyproduct{500x500}{\micro\meter}, consisting of about 150 cells. The maximum glucose import $U_\gluc$ and maximum oxygen import $U_\ox$ were chosen to be \qty{1}{\mmolgh} and \qty{3}{\mmolgh} respectively. The ATP maintenance demand $L_\maint$ was fixed at \qty{1}{\mmolgh}. All parameters for the model and simulations obtained from literature, including the diffusion coefficients used for the relevant metabolites, are provided in \suptable{1}, along with other relevant derived quantities.

\subsection{The sampling algorithm}

The space of feasible multi-cellular metabolic flux configurations  $\mathcal{F}_N$ was sampled according to the specified probability distribution \eqref{eq:boltz2} using a hit-and-run Markov chain algorithm. The steps are provided in detail in \supmethod{2}. The code to perform the sampling is made available online \cite{https://doi.org/10.5281/zenodo.12771735}.

\subsection{Mean-field approximation}

The detailed derivation of the explicit partition function in terms of the parameters is provided in \supmethod{3}. See also \supmethod{4} for details on how the mean-field equations are modified when a background lactate term is introduced. The code to calculate various quantities under the mean-field approximation is made available online \cite{https://doi.org/10.5281/zenodo.12771735}.

\subsection{Inverse modeling experimental data}

For inference of the model parameters that yield the best prediction of time-resolved experimental lactate flux data \cite{onesto2022probing, https://doi.org/10.5281/zenodo.12772220}, predictions from the model were fitted to key statistical features of the observed flux distribution: the average lactate flux, its standard deviation (fluctuations), and nearest-neighbor correlations. The model parameters $\beta_\gluc$ and $\beta_\ox$ were allowed to vary for each experimental time point, and the phenomenological nearest-neighbor coupling constant was kept constant. The inference procedure involved maximizing the likelihood of experimental data and sampling parameters according to the posterior probability distribution using a Metropolis Monte-Carlo algorithm, and is described in detail in \supmethod{8}, along with the various attempts that were made including and excluding different terms in the likelihood function before the final form was used.

\section{Data availability}

\added{The experimental data analyzed in this work were published previously \cite{onesto2022probing} and are available online \cite{https://doi.org/10.5281/zenodo.12772220}.}

\section{Code availability}

\added{The code to perform simulations, mean field calculations and inference are available online \cite{https://doi.org/10.5281/zenodo.12771735}.}

\bibliography{main}

\backmatter

\bmhead{Acknowledgements}

KN and DDM thank the FBB (Fundación Biofísica Bizkaia) for support.
\added{DDM acknowledges financial support from the grants PIBA\_2024\_1\_0016 (Basque Government) and PID2023\_146408NB\_I00 (Spanish Ministry of University and Reserach).}
JAPM, RM and ADM acknowledge financial support from the European REA, Marie Sk{\l}odowska-Curie Actions, grant agreement no. 101131463 (SIMBAD).
DST thanks Biofisika Institute for hosting him while part of this work was carried out.
LLDM thanks the Associazione Italiana per la Ricerca contro il Cancro (AIRC) (MFAG-2019, n. 22902), \deleted{and} the PRIN 2022 (2022CRFNCP) funded by the Italian Ministry of Research (MUR) European Union – Next Generation EU \added{and the Italian Ministry of Research, under the complementary actions to the NRRP `Fit4MedRob - Fit for Medical Robotics' Grant (\#PNC0000007).}

\bmhead{Author contributions}
Conceptualization: DDM, DT, ADM, RM, and LDM.
Modeling: KN, JAPM, AF, DDM, DT, ADM and RM.
Simulations: KN, JAPM and DDM.
Analytical calculations: KN, DDM, DT and ADM.
Data Analysis and Inference: KN, JAPM, AF, VO and DDM.
Experiments: VO, SF and LDM.
Funding: DDM, DT, ADM, RM, and LDM.

\bmhead{Competing interests}
The authors declare no competing interests.

\bmhead{Supplementary information}
The online version is accompanied by supplementary material containing further information on the constraint-based model parameters, experimental data, mean field calculations and the inference pipeline.

\end{document}


\maketitle

\tableofcontents


\mypart{Supplementary Notes}
\renewcommand{\sectype}{NOTE}

\section{Model parameters and constants}

In Supplementary Table \ref{table:parameters}, we provide the actual values of the parameters and constants used with the metabolic network model, including the inter-cellular diffusion constraints.


\begin{table}[ht]
\setlength\extrarowheight{0.6mm}
\centering
\resizebox{\textwidth}{!}{
\begin{tabular}{lccl}
    \hline
        \multicolumn{1}{c}{Name (ID)} &
        Value/range &
        Unit &
        Reference \\
    \hline
        Cell diameter ($R$) &
        10  &   \si{\micro\meter} &
        This work
        \\
        Linear size of culture ($L$)&
        1   &   \si{\centi\meter} &
        This work
        \\
        Number of cells ($N$)&
        $4 \times 10^4$ & cells &
        This work
        \\
        Mean field constant ($K$)&
        40 & unit-less &
        This work**
        \\
        Average cell dry weight ($m$) &
        1   & \si{\nano\gram} &
        \cite{amMeanMassspecificMetabolic2008}
        \\
         Maximum molecular oxygen import  ($U_\ox$) &
        $3$   &   \si{\mmolgh}    &
        \cite{fernandez-de-cossio-diazCharacterizingSteadyStates2017, wagnerRateOxygenUtilization2011}
        \\
        Maximum glucose import ($U_\gluc$) &
        1     &   \si{\mmolgh}    &
        \cite{fernandez-de-cossio-diazCharacterizingSteadyStates2017, szeliovaInclusionMaintenanceEnergy2021}
        \\
        ATP maintenance demand ($L_\maint$) &
        $1$  &   \si{\mmolgh}    &
        \cite{fernandez-de-cossio-diazCharacterizingSteadyStates2017, szeliovaInclusionMaintenanceEnergy2021}
        \\
        ATP produced by fermentation &
        1 & ATP per half glucose &
        \cite{nelsonLehningerPrinciplesBiochemistry2013}
        \\
        ATP produced by respiration &
        5 & ATP per oxygen (\ce{O2}) &
        \cite{nelsonLehningerPrinciplesBiochemistry2013}
        \\
        Glucose medium concentrations ($c_\gluc$) &
        25 & \si{\milli\molar} &
        This work
        \\
        Oxygen medium concentrations ($c_\ox$) &
        0.25 & \si{\milli\molar} &
        This work
        \\
        Glucose diffusion constants ($D_\gluc$) &
        600     & \si{\micro\meter\squared\per\second} &
        \cite{hoberPhysicalChemistryCells1946}
        \\
        Oxygen diffusion constants ($D_\ox$) &
        2000    & \si{\micro\meter\squared\per\second} &
        \cite{subczynskiDiffusionOxygenWater1984}
        \\
        Lactate diffusion constants ($D_\lact$) &
        700     & \si{\micro\meter\squared\per\second} &
        \cite{hoberPhysicalChemistryCells1946}
        \\
        Proton diffusion constants ($D_{\ce{H+}}$) &
        7000    & \si{\micro\meter\squared\per\second} &
        \cite{hoberPhysicalChemistryCells1946}
        \\
        Oxygen diffusion derived maximum intake ($u_\ox^\text{max}$) &
        230 &   \si{\mmolgh}    &
        This work*
        \\
        Glucose diffusion derived maximum intake ($u_\gluc^\text{max}$)&
        6800    &   \si{\mmolgh}    &
        This work*
        \\
    \hline
    \vspace{-10pt}
\end{tabular}
}
    (*) $u_{\textnormal{\textsc{x}}}^\text{max} = 4 \pi c_{\textnormal{\textsc{x}}} D_{\textnormal{\textsc{x}}} R/m$
    \qquad
    (**) $K =  N R / L$
\caption{
    Relevant constants and parameters of the multicellular metabolic network model.
}
\label{table:parameters}
\end{table}

\section{Multi-cellular diffusion constraints}


Consider a group of $N$ spherical cells of identical radius $R$, with centers  placed at positions $\rr_i$ in a three-dimensional volume $V$ ($i=1,\ldots,N$), such that $|\rr_i-\rr_j|>  2R$ for any $i\neq j$. We assume that each cell is either a net absorber (flux $u_i>0$) or a net emitter (flux $u_i<0$) of a certain compound 
whose extracellular concentration is represented by a scalar field $c\equiv c(\rr,t|\rr_c,\mathbf{u})$, where $\rr_c=\{\rr_i\}$ and $\mathbf{u}=\{u_i\}$. Assuming that no material flow is present in the extracellular fluid and that neither the positions $\rr_c$ of cells nor $\mathbf{u}$ change in time, such a field evolves due to (i) cellular emission and absorption, and (ii) random diffusion in $V$, leading to the diffusive problem
\begin{equation}\label{diffu}
\frac{\partial c}{\partial t}=D\nabla^2 c~~.
\end{equation}
complemented by suited boundary conditions on the cells surfaces.

We are interested in characterizing its steady state, i.e. the solutions of the {\it Laplace equation} \begin{gather}\label{lapl}
\nabla^2 c=0~~.
\end{gather}

Basic mathematical properties like existence and uniqueness of solutions for different classes of boundary conditions are presented in great detail in classical physics textbooks such as \cite{griffiths2021introduction}, Chapters 2 and 3. We focus here on features that are specifically important for the present work.

To begin with, we recall that, for a perfect, isolated spherical absorber with $c=0$ on the surface and $c=c_\infty$ for $|\rr|\equiv r\to\infty$, the time-dependent diffusion equation (\ref{diffu}) is solved by the radial function
\begin{equation}
c(r,t)=c_\infty\left[1-\frac{R}{r}\left(1-\text{erf}\,\frac{r-R}{\sqrt{4Dt}}\right)\right]~~.
\end{equation}
Correspondingly, using the diffusion flux $\mathbf{j}=-D\nabla c$, the cellular intake rate is given by
\begin{equation}
u(t)=4\pi R^2 |j(R,t)|=4\pi D R c_\infty\left(1+\frac{R}{\sqrt{\pi D t}}\right)~~.
\end{equation}
For $t\to\infty$ one gets
\begin{equation}
u\to u_s\equiv 4\pi D R c_\infty~~~\text{and}~~~ c\to c_\infty\left(1-\frac{R}{r}\right)=c_\infty-\frac{u_s}{4\pi D r}~~.
\end{equation}
(The formula for $u_s$ is known as Smoluchowski's formula.) In the presence of multiple absorbers, shielding effects become relevant and $u_s$ only represents an upper bound to the intake rate of an absorber. Because of the linearity of Laplace's equation, the steady state solution with $N$ absorbers reads
\begin{equation}\label{foweuh}
c(r)=c_\infty-\sum_i\frac{u_i}{4\pi D |\rr-\rr_i|}~~,
\end{equation}
where $u_i$ represents the uptake flux of cell $i$. For the sake of stability, at the position of cell $j$ we must have
\begin{equation}
c_\infty\geq \sum_i\frac{u_i}{4\pi D |\rr_j-\rr_i|}~~,
\end{equation}
or, equivalently,
\begin{equation}\label{dc}
u_s\geq \sum_i A_{ji}u_i~~~~~(\forall j)~~,
\end{equation}
where
\begin{equation}
A_{ji}=\delta_{ji}+(1-\delta_{ji})\frac{R}{|\rr_j-\rr_i|}~~.
\end{equation}
(We used Kronecker's $\delta$-symbol: $\delta_{ij}=1$ if $i=j$, and zero otherwise.) Eq. (\ref{dc}) can be seen as a global constraint that diffusion imposes on the feasible values of fluxes $u_i$. Notice that expression (\ref{foweuh}) holds in principle also in the presence of emitters with $u_i<0$. Finally, if there is no reservoir of particles at infinity (i.e. if $c_\infty=0$), then (\ref{dc}) becomes
\begin{equation}
\sum_i A_{ji}u_i\leq 0~~~~~(\forall j)~~.
\end{equation}





\section{Inter-cellular exchanges and overflow}
Consider a vector $\mathbf{u} = (u_1,..., u_N) \in \mathbb{R}^N$ and invertible square matrix $\mathbf{A} \in \mathbb{R}^{N \times N}$. In the main text these are identified with the vector of lactate flux values (i.e. $u_i= - u^{(i)}_\lact$ is the flux of lactate for the $i^\text{th}$ cell, note the change in sign) and diffusion constraint matrix $\mathbf{A} $. Let $\mathbf{1} = (1,...,1)^T$ be the $N$-dimensional vector of all ones and $\mathbf{e}_i$ be the $N$-dimensional unit coordinate vector with one in the $i^\text{th}$ position and zero everywhere else. Then, if we define the extended matrix
\begin{equation}
\mathbf{M} = \begin{pmatrix} \mathbf{A} & \mathbf{1} & -\mathbf{1}  \end{pmatrix}
\end{equation}
we have the following equivalence of constraints
\begin{equation}
\label{diffusionconstraints}
\mathbf{u}^T \cdot \mathbf{M} \geq 0 \quad \iff \quad  \mathbf{u}^T \cdot \mathbf{A} \geq 0 , \quad \mathbf{u}^T \cdot \mathbf{1} =  \sum_{i=1}^N u_i = 0 .
\end{equation}
In terms of the main text, this is the statement that a set of lactate flux values satisfies the diffusion constraints with the additional condition that there is no net lactate production. In particular, apart from certain contrived cases, a Flux Balance Analysis optimal yield flux pattern (where all cells are at point \textsf{E} in Figure 1b from the main text) must have this solution form, since otherwise it would be considered sub-optimal (i.e., there would always be another solution that could obtain a higher value of the objective function by decreasing net lactate production). We now derive a general condition on $\mathbf{A}$ that guarantees the only solution to these constraints is the trivial solution $u_1 = ... = u_N = 0$.

First note that, for any non-trivial solution $\mathbf{u}$, there exists some $i$ such that $u_i =\mathbf{u}^T \cdot \mathbf{e}_i  <0$ without loss of generality (since from the constraint $\mathbf{u}^T \cdot \mathbf{1} =0$ then we must also have some $j \neq i$ with $u_j >0$). We can therefore write the conditions on a  non-trivial solution as
 \begin{equation}
\mathbf{u}^T \cdot \mathbf{M} \geq 0,  \quad \mathbf{u}^T \cdot  \mathbf{e}_i  <0 ,
\end{equation}
and from Farkas' Lemma we can have no such $\mathbf{u}$ if we can find a solution to the system of equations
 \begin{equation}
  \mathbf{M}  \cdot \mathbf{y} = \mathbf{e}_i, \quad \mathbf{y} \geq 0 .
\end{equation}
Here $\mathbf{y} \geq 0$ means that all components of $\mathbf{y}$ are non-negative. Explicitly, we have
 \begin{equation}
 \begin{pmatrix} \mathbf{A} & \mathbf{1} & -\mathbf{1}  \end{pmatrix}  \begin{pmatrix} \mathbf{x} \\ y_2 \\ y_3  \end{pmatrix}=  \mathbf{A} \cdot \mathbf{x} + (y_2 - y_3) \mathbf{1} = \mathbf{e}_i ,
\end{equation}
where $\mathbf{x} \geq 0$ and from now on we identify $\lambda \equiv y_2 - y_3 \in \mathbb{R}$. Since $\mathbf{A}$ is invertible, we have
\begin{equation}
\mathbf{x} = \mathbf{A} ^{-1} \cdot( \mathbf{e}_i- \lambda  \mathbf{1})
\end{equation}
and
\begin{equation}
x_j = \mathbf{e}_j^T \cdot \mathbf{x} = (\mathbf{A}^{-1})_{ji} - \lambda \sum_{k=1}^N (\mathbf{A}^{-1})_{jk}  \equiv a_{ji} - \lambda B_j
\end{equation}
where $a_{ji}$ is the $(i,j)^\text{th}$ component of the inverse matrix $\mathbf{A}^{-1}$ and $B_j$ is the sum over elements in the $j^\text{th}$ column of $\mathbf{A}^{-1}$. Provided $B_j$ is non-zero, from the condition $x_j \geq 0 $ for all $j$ we get that we can always find $\lambda \in \mathbb{R}$ whenever
\begin{equation}
\min_{j: B_j >0} \frac{a_{ji}}{B_j} \geq \max_{j: B_j <0} \frac{a_{ji}}{B_j}.
\end{equation}
It therefore follows that, provided this condition on $\mathbf{A}$ is satisfied for all $i$, then there will be no non-trivial solution to the conditions (\ref{diffusionconstraints}). It can be checked numerically that this condition is satisfied for any sensible choice of inter-cell distance matrix and geometrically corresponds to the intersection of all separating hyper-planes defined by the constraints meeting at just one point. Specifically, this implies that there will be no optimal solution to the maximum yield Flux Balance Analysis problem with lactate exchange between cells.

\mypart{Supplementary Methods}
\renewcommand{\sectype}{METHOD}

\section{The partition function}
\label{sec:partfunc}

The partition function $Z_N$
for a system of $N$ cells is a function of Lagrange multipliers $\beta_\gluc$ and $\beta_\ox$ given by the following $2N$-dimensional integral over the multi-cellular flux space $\mathcal{F}_N$:
\begin{equation}
    Z_{N}
        \left( \beta_\gluc, \beta_\ox \right)
    =      \int_{\mathcal{F}_N}
            \prod_{n=1}^N
            du_\gluc^{(n)} \, du_\ox^{(n)} \,
            e^{\beta_\gluc u_\gluc^{(n)}} e^{\beta_\ox u_\ox^{(n)}} \;
            .
    \label{eq:Z_N_rough}
\end{equation}
%


%

The partition function also serves as a moment-generating function for the expectation values and higher moments of net fluxes (normalised by the number of cells) defined with respect to the maximum entropy probability measure over multi-cellular flux space. Specifically, given the exponential form of the integrand in (\ref{eq:Z_N_rough}), we have that
\begin{equation}
    \expect{u_\gluc}
        =   \frac{1}{N} \der{}{\beta_\gluc} \ln{Z_N}(\beta_\gluc,\beta_\ox)
    \quad \text{;} \quad
    \expect{u_\ox}
        =   \frac{1}{N}  \der{}{\beta_\ox} \ln{Z_N}(\beta_\gluc,\beta_\ox)   ,
\end{equation}
and, by stoichiometry,
\begin{equation}
    \expect{u_\lact} = \frac{\expect{u_\ox}}{3} - 2 \expect{u_\gluc}   .
    \label{eq:mean-lactate-full}
\end{equation}

Similarly, variances of these flux values are given by the second-order partial derivatives of $Z_N$
\begin{equation}
    \sigma^2_{u_\gluc} \equiv \expect{u_\gluc^2} - \expect{u_\gluc}^2
        =   \frac{1}{N^2}\dder{}{\beta_\gluc} \ln{Z_N}(\beta_\gluc,\beta_\ox)
    \quad \text{;} \quad
    \sigma^2_{u_\ox} \equiv \expect{u_\ox^2} - \expect{u_\ox}^2
        =   \frac{1}{N^2}\dder{}{\beta_\ox} \ln{Z_N}(\beta_\gluc,\beta_\ox)   .
\end{equation}

For the variance of the net lactate flux, we use that
\begin{alignat}{3}
    \expect{u_\lact^2} &= \expect{\left(\frac{u_\ox}{3} - 2u_\gluc\right)^2}%
                &&= \frac{\expect{u_\ox^2}}{9} + 4\expect{u_\gluc^2} - \frac{4}{3} \expect{u_\ox \, u_\gluc}   &  \\
    \expect{u_\lact}^2 &= \left( \frac{\expect{u_\ox}}{3} - 2 \expect{u_\gluc} \right) ^2%
                &&= \frac{\expect{u_\ox}^2}{9} + 4\expect{u_\gluc}^2 - \frac{4}{3} \expect{u_\ox} \expect{u_\gluc} &
\end{alignat}
which gives
\begin{equation}
    \sigma_{u_\lact}^2  =   \left( \frac{\sigma_{u_\ox}}{3} \right) ^2
                        +   \left( - 2 \sigma_{u_\gluc} \right) ^2
                        -   \frac{4}{3} \Corr{u_\ox,u_\gluc}
    \label{eq:variance-lactate}
\end{equation}
where
\begin{equation}
    \Corr{u_\ox,u_\gluc}  \equiv  \expect{u_\ox \, u_\gluc} - \expect{u_\ox} \expect{u_\gluc}
    =  \frac{1}{N^2}\dderd{}{\beta_\gluc}{\beta_\ox} \ln{Z_N}(\beta_\gluc,\beta_\ox) .
\end{equation}

Thus, once $Z_{N} \left( \beta_\gluc, \beta_\ox \right)$ is known, one can obtain the expectation values and variances of any net flux value as a function of $\beta_\gluc$ and $\beta_\ox$.

We highlight that, in a more general setting, it is of course possible to introduce $2N$ Lagrange multipliers, one for each cell, that constrain expectation values of each single-cell flux value individually. In that case, the partition function would depend on $2N$ parameters and individual single-cell moments of the flux distribution obtained by differentiation in the same way as described for net flux values above. However, in this work we restrict our focus to a simple, two-parameter model with the goal of parsimoniously reproducing statistical features of existing experimental data and describing the collective behavior of multi-cellular flux distributions. More complex models of the same nature could be applied to experimental data that becomes available in the future.

\section{The sampling algorithm \label{sec:sampling}}

The multi-cellular metabolic model defines a high-dimensional convex polytope ($D\,{=}\,2N\,{\approx}\,300$, where $N$ is the number of cells) and the computational task at hand is to characterize this space, in this case with flux distributions weighted by a Boltzmann factor defined by the maximum entropy constraint. This problem is connected to a class of NP-hard computational problems such as the computation of a matrix permanent, volume of a high-dimensional convex body or the Ising partition function, which can be solved numerically in polynomial time using Markov chain Monte Carlo algorithms \cite{simonovits2003compute}. In particular, over-relaxed algorithms like hit-and-run \cite{turcin1971computation} have been shown to work very well in the context of constraint-based genome-scale metabolic modeling, once the ill-conditioning problems connected to heterogeneous scales have been tackled by approximate ellipsoidal rounding \cite{de2015uniform}. In this work we employed an hit-and-run Markov chain to sample the space, outlined schematically by the following workflow:
\begin{enumerate}
\setcounter{enumi}{-1}
\item Initial data: a point $P_0$ inside the polytope (can be found with a relaxation algorithm such as \cite{krauth1987learning}).
\item Given point $P_i$, generate a direction/versor $\hat{n}$ uniformly at random (a point on a unit hyper-sphere, e.g. with the Marsaglia method \cite{knuth2014art}).
\item Find the intersections $t_1,t_2$ of the line $L(t) = P_i + t \hat{n}, \quad t\in \mathbb{R}$ with the boundary of the polytope.
\item Extract $t^* \in [t_1,t_2]$ by inverting the cumulative distribution function  of the marginalized Boltzmann distribution over the segment \cite{knuth2014art}. Set $P_{i+1} = P_i + t^* \hat{n}$. Return to step 1 (or end the algorithm if you think you have enough points).
\end{enumerate}
It is interesting to notice that, by comparison with the case of sampling steady states of bulk genome scale metabolic  networks, the ill-conditioning problem is much less severe in our case. This is due the symmetric structure of the space, which
is given by the product of $N$ identical single-cell metabolic flux spaces (plus  the diffusion constraints). A code implementation of the sampling algorithm  is provided in the repository \url{https://github.com/KrishnadevN/MulticellularMetabolicNetworks}.

\section{The mean-field approximation}

\subsection{Approximation of the partition function}
The full partition function \eqref{eq:Z_N_rough} can be written explicitly as
\begin{equation*}
    Z_N(\beta_\gluc,\beta_\ox)
        =   \int_0^{U_\gluc}    \;
            \int_0^{U_\ox}
            \cdots
            \int_0^{U_\gluc}    \;
            \int_0^{U_\ox}    \;
            \prod_{n=1}^N
            du_\gluc^{(n)} \, du_\ox^{(n)} \,
            e^{\beta_\gluc u_\gluc^{(n)}+ \beta_\ox u_\ox^{(n)}} \;
            \theta \big( f_\atp^{(n)} - L_\maint \big)    \,
            \theta \biggl(  - u_\lact^{(n)}
                            - \sum_{\substack{m = 1 \\ m \neq n}}^N u_\lact^{(m)} A_{mn} \biggr)
\end{equation*}
where we have used the Heaviside step function
\begin{equation*}
    \theta (x)  \equiv  \begin{cases}
                            1,  & x \ge 0   \\
                            0,  & x <   0,
                        \end{cases}
\end{equation*}
to impose the constraints of minimal ATP demand and those given by the diffusion of lactate in the medium. Here, $u_\lact^{(n)}$ and $f_\atp^{(n)}$ are related to the independent variables $u_\gluc^{(n)}$ and $u_\ox^{(n)}$ as
\begin{equation}
    u_\lact^{(n)} = \frac{u_\ox^{(n)}}{3} - 2 u_\gluc^{(n)}
    \quad \text{and} \quad
    f_\atp^{(n)} = 2 u_\gluc^{(n)} + \frac{14}{3} u_\ox^{(n)}.
\end{equation}

To obtain an exact expression for $Z_N$, we make the approximation that $A_{mn} \approx K/N$
for all cells $m \neq n$, which is equivalent to the mean-field approximation that all cells are fully connected and equally distanced, as discussed in the main text. It is important to remark that this approximation also assumes the system to be homogeneous, whereas in general the partition function could depend on the specific spatial positions of individual cells. This homogeneity assumption
also means that the moments of the single-cell distribution do not depend on the cell index $(n)$, i.e. the fluxes are identically distributed  (albeit correlated) random variables. In particular, when $N\gg 1$ we have
\begin{equation}
    \frac{1}{N} \sum_{\substack{m = 1 \\ m \neq n}}^N u_\lact^{(m)}
    \approx
    \frac{1}{N} \sum_{m=1}^N u_\lact^{(m)}
    \equiv
    \avgul ,
\end{equation}
i.e., the contribution of any one cell to the arithmetic mean across cells, $\avgul$, becomes negligible. With this approximations, the inter-cellular constraint originating from the diffusion of lactate becomes
\begin{equation}
                \theta \biggl( - u_\lact^{(n)} - \sum_{\substack{m = 1 \\ m \neq n}}^N
                                                    u_\lact^{(m)} A_{mn}      \biggr)
        \approx \theta \Bigl(  - u_\lact^{(n)} - K \, \avgul  \Bigr)
\end{equation}
so that
\begin{equation}
    Z_N(\beta_\gluc,\beta_\ox)  
        \approx \int_0^{U_\gluc}    \;
                \int_0^{U_\ox}
                \cdots
                \int_0^{U_\gluc}    \;
                \int_0^{U_\ox}    \;
                \prod_{n=1}^N
                du_\gluc^{(n)} \, du_\ox^{(n)} \,
                e^{\beta_\gluc u_\gluc^{(n)}+ \beta_\ox u_\ox^{(n)}} \;
                \theta \bigl( f_\text{ATP}^{(n)} - L_\maint \bigr)   \;
                \theta \bigl( - u_\lact^{(n)} - K \avgul \bigr)  .
                \label{eq:approx_Z}
\end{equation}
%

To manipulate the above integral into a manageable form, we introduce $\phi$ and $\lambda$ through the properties of the Dirac delta and its Fourier transform, to express unity as
\begin{align}
    1   &=  \;
            N \int_{-\infty}^\infty
                \delta \left( N \phi  - \sum_{n=1}^N u_\lact^{(n)} \right) d \phi      \\
        &=  \; \frac{N}{2 \pi}
            \int_{-\infty}^\infty
            \int_{-\infty}^\infty
                    \exp \left( \I \lambda \left[ N \phi
                                            - \sum_{n=1}^N u_\lact^{(n)} \right]\right)
                                            d \phi \, d \lambda        \\
        &=  \; \frac{N}{2 \pi}
            \int_{-\infty}^\infty
                d\phi
            \int_{-\infty}^\infty
                d\lambda \,
                    e^{ i \lambda N \phi }
                    \prod_{n=1}^N
                    e^{ 2 i \lambda u_\gluc^{(n)} - i \lambda u_\ox^{(n)}/3 }.
        \label{eq:unity}
\end{align}
Inserting this into the integrand of (\ref{eq:approx_Z}), we obtain the identity
\begin{equation*}
                \prod_{n=1}^N e^{\beta_\gluc u_\gluc^{(n)}+ \beta_\ox u_\ox^{(n)}} \theta \bigl( - u_\lact^{(n)} - K \avgul \bigr) = \frac{N}{2 \pi}
            \int_{-\infty}^\infty
                d\phi
            \int_{-\infty}^\infty
                d\lambda \,
                    e^{ i \lambda N \phi } \prod_{n=1}^N e^{(\beta_\gluc + 2 i \lambda) u_\gluc^{(n)} + (\beta_\ox - i \lambda/3) u_\ox^{(n)}}\theta \biggl(  - u_\lact^{(n)}
                            - K \phi  \biggr),
\end{equation*}
which enables the approximation of the partition function (\ref{eq:approx_Z}) to be written as
\begin{align}
    Z_N \left( \beta_\gluc, \beta_\ox \right)  
        &\approx
           \; \frac{N}{2 \pi}
            \int_{-\infty}^\infty
                d\phi
            \int_{-\infty}^\infty
                d\lambda \,
                    e^{ i \lambda N \phi }
                    [Z   \bigl(  \beta_\gluc + 2 i \lambda, \,
                                    \beta_\ox - i \lambda/3
                                    , \,
                                    \phi            \bigr)  ]^N
        \label{eq:decoupled}                                            \\
        &=  \frac{1}{2 \pi}
            \int_{-\infty}^\infty
                d\phi
            \int_{-\infty}^{\infty}
                d\lambda \,
                    \exp{   \left[  N F(\beta_\gluc, \beta_\ox ,\phi, i \lambda )
                            \right] }
        \label{eq:Z_N_approx}
\end{align}
with
\begin{equation}
    Z(\beta_\gluc,\beta_\ox, \phi) =
            \int_{0}^{U_\gluc}
            \int_{0}^{U_\ox} du_\gluc du_\ox \,
            e^{\beta_\gluc u_\gluc} e^{\beta_\ox u_\ox} \;
                \theta \bigl( f_\text{ATP} - L_\maint \bigr)   \;
                \theta \bigl( - u_\lact - K \phi \bigr)
        \label{eq:factor-explicit}
\end{equation}
and
\begin{equation}
    F(\beta_\gluc, \beta_\ox ,\phi, p )
        =  p \phi
                + \log Z  \bigl(  \beta_\gluc + 2 p,
                                    \beta_\ox - p/3
                                    ,   \,
                                    \phi        \bigr)
                + \frac{1}{N} \log N  .
    \label{eq:F}
\end{equation}

%
%
The introduction of $\phi$ and $\lambda$ has thus allowed us to ``decouple'' the diffusion constraints to study the effective two-dimensional, single-cell flux space on which $Z$ is defined.

%
%
%
%
%
Finally, to evaluate \eqref{eq:Z_N_approx} in the limit $N \to \infty$, where the $\log(N)/N$ term in (\ref{eq:F}) can safely be ignored, we use the method
of steepest descent (also called the saddle point approximation) \cite{bender2013advanced}. For $N \gg 1$, we obtain
\begin{equation}
    \frac{1}{N} \log Z_N \left( \beta_\gluc, \beta_\ox \right)
    \; \approx \;
    p_* \phi_*
        + \log Z  \bigl(  \beta_\gluc + 2 p_*,
                            \beta_\ox - p_*/3
                            ,   \,
                            \phi_*  \bigr)
    \label{eq:Z_N_ident}
\end{equation}
where $\left( \phi_*, p_* \equiv i\lambda^*\right)$ is the stationary point given by
\begin{equation}
    \left. \der{F}{\phi}  \right|_{\left( \phi_*, p_* \right)} = 0
    \qquad \text{and} \qquad
    \left. \der{F}{p} \right|_{\left( \phi_*, p_* \right)} = 0  .
\end{equation}
In particular, from \eqref{eq:F}, these saddle point equations correspond to the self-consistency equations
\begin{align}
    p^*(\beta_\gluc,\beta_\ox) &=    - \der{}{\phi} \ln{Z}
                           \bigl(  \beta_\gluc + 2 p_*,
                            \beta_\ox - p_*/3
                            ,   \,
                            \phi_*  \bigr)
                            \label{eq:pstar}
                             \\
    \phi^*(\beta_\gluc,\beta_\ox)  &=    - \der{}{p} \ln{Z}
                            \bigl(  \beta_\gluc + 2 p_*,
                            \beta_\ox - p_*/3
                            ,   \,
                            \phi_*  \bigr)
    \label{eq:ulstar}
\end{align}
where the explicit dependence of $p^*$ and $\phi^*$ on $\beta_\gluc$ and $\beta_\ox$ is indicated. In fact, by using the chain rule we see that
\begin{equation}
    \phi^*(\beta_\gluc,\beta_\ox) = \left( \frac{1}{3}\der{}{\beta_\ox} - 2\der{}{\beta_\gluc} \right) \log Z  \bigl(  \beta_\gluc + 2 p_*,
                            \beta_\ox - p_*/3
                            ,   \,
                            \phi_*  \bigr) \equiv \newexpect{u_\lact}
                    \label{eq:phistar}
\end{equation}
where $\newexpect{\cdots}$ is the expectation value defined using the measure associated with $Z$ on the effective single-cell space. Namely, from (\ref{eq:factor-explicit}) and (\ref{eq:phistar}) with $\beta_\gluc^* \equiv \beta_\gluc +2p^*$ and $\beta_\ox^* = \beta_\ox -p^*/3$ we have, self-consistently,
\begin{equation}
    \newexpect{\cdots} \equiv \frac{1}{Z  (  \beta_\gluc^*,
                            \beta_\ox^*,
                            \newexpect{u_\lact})}
        \int_{0}^{U_\gluc}
            \int_{0}^{U_\ox} du_\gluc du_\ox \,
            (\cdots)e^{\beta_\gluc^* u_\gluc} e^{\beta_\ox^* u_\ox} \;
                \theta \bigl( f_\text{ATP} - L_\maint \bigr)   \;
                \theta \bigl( - u_\lact - K \newexpect{u_\lact} \bigr)
                \label{eq:mfmeasure}
        .
\end{equation}
The values $\phi^*$ and $p^*$ therefore permit approximation of $Z_N$ using (\ref{eq:Z_N_ident}) once $Z$ and its derivatives are known. We provide exact analytical formulae for these in the next subsection.

\subsection{\label{sec:single-cell-Z}Analytical form of $Z(\beta_\gluc,\beta_\ox, \phi)$ and its derivatives }


We now proceed to evaluate the integral defined in (\ref{eq:factor-explicit}) to find an analytical formula for $Z(\beta_\gluc,\beta_\ox, \phi)$ and its derivatives, which are required for computation of the full partition function $Z_N$. We recall that $Z$ is defined on the effective single-cell space illustrated in Supplementary Figure \ref{fig:flux-space-meanfield}, restricted to the domain $\phi \leq 0$ due to its evaluation at $\phi^*$, as described in the previous subsection. However, we remind the reader that the self-consistent dependence of $\phi^*$ on $\beta_\gluc$ and $\beta_\ox$ is only imposed after the saddle point approximation so that here $\phi$ is treated as an independent argument of the three-variable function $Z(\beta_\gluc,\beta_\ox, \phi)$. The resulting domain of integration can be intuitively thought of as the single-cell flux space from the main text (Figure 1b) combined with an additional upper bound on the rate of lactate uptake, given by the constraint $u_\lact \leq - K \phi$.

\begin{figure}[htb]
    \centering
    \includegraphics[trim={0 0.5cm 0 0.5cm}, clip]{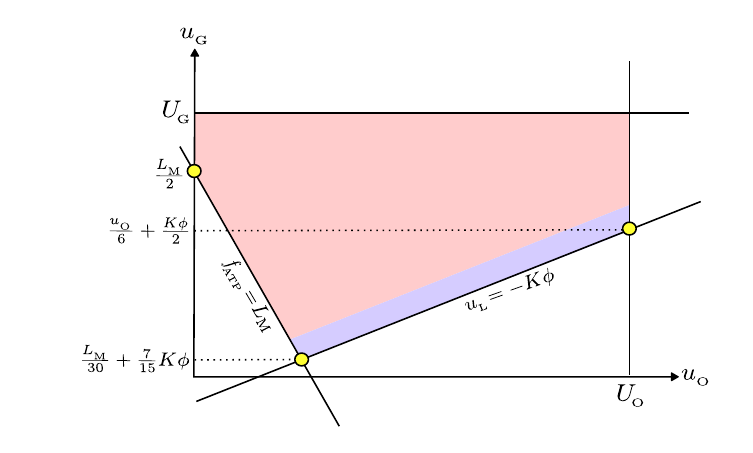}
    \caption{The effective single-cell flux space (in color) given by the limits of integration for $Z(\beta_\gluc,\beta_\ox, \phi)$. Analogously to the single-cell flux space described in the main text, the pink region has the physical interpretation of lactate export and the purple region denotes lactate import.}
    \label{fig:flux-space-meanfield}
\end{figure}

In terms of variables $u_\gluc,u_\ox$, these constraints give the following limits of integration for the integral in (\ref{eq:factor-explicit}):
\begin{equation}
    u_\gluc^\text{min}      \le     u_\gluc     \le     U_\gluc
    \qquad ; \qquad
    u_\ox^\text{min}      \le     u_\ox     \le     u_\ox^\text{max}
\end{equation}
where,
\begin{equation}
\begin{split}
    u_\gluc^\text{min}  &=  \max    \left( 0,\; \frac{L_\maint}{30} + \frac{7}{15} K \phi
                                \right) \\
    u_\ox^\text{min}  &=  \begin{cases}
                            \frac{3}{14} \left( L_\maint - 2u_\gluc \right),
                                &   u_\gluc^\text{min}  \le     u_\gluc     \le     u_\gluc^a   \\
                            0,
                                &   u_\gluc^a           \le     u_\gluc     \le     U_\gluc
                        \end{cases}
                        ; \quad     u_\gluc^a   \equiv
                                    \min \left( \frac{L_\maint}{2}, \; U_\gluc \right) \\
    u_\ox^\text{max}  &=  \begin{cases}
                            6 u_\gluc -3 K \phi,
                                &   u_\gluc^\text{min}  \le     u_\gluc     \le     u_\gluc^b   \\
                            U_\ox,
                                &   u_\gluc^b           \le     u_\gluc     \le     U_\gluc
                        \end{cases}
                        ; \quad     u_\gluc^b   \equiv
                                    \max    \left( 0, \;
                                    \frac{U_\ox}{6} + \frac{K \phi}{2}
                                            \right).
\end{split}
\label{eq:bounds}
\end{equation}
We evaluate
\begin{align}
    Z( \beta_\gluc, \beta_\ox ,\phi)
        &=  \int_{u_\gluc^\text{min}}^{U_\gluc} du_\gluc
            \int_{u_\ox^\text{min}}^{u_\ox^\text{max}} du_\ox
                \, e^{\beta_\gluc u_\gluc} e^{\beta_\ox u_\ox}                  \\
        &=  \frac{1}{\beta_\ox}   \int_{u_\gluc^\text{min}}^{U_\gluc} du_\gluc
                \, e^{\beta_\gluc u_\gluc} e^{\beta_\ox u_\ox^\text{max}}
        -   \frac{1}{\beta_\ox}   \int_{u_\gluc^\text{min}}^{U_\gluc} du_\gluc
                \, e^{\beta_\gluc u_\gluc} e^{\beta_\ox u_\ox^\text{min}}.
\end{align}

Since the dependence of $u_\ox^\text{min}$ and $u_\ox^\text{max}$ on $u_\gluc$ changes at $u_\gluc^a$ and $u_\gluc^b$, respectively, we split each integral into two and obtain
\begin{align}
\begin{split}
    Z( \beta_\gluc, \beta_\ox ,\phi)   \;=\;
        &   \frac{1}{\beta_\ox}   \int_{u_\gluc^\text{min}}^{u_\gluc^b} du_\gluc
                \, e^{(\beta_\gluc + 6 \beta_\ox) u_\gluc} e^{-3 K \phi \beta_\ox}
        +   \frac{1}{\beta_\ox}   \int_{u_\gluc^b}^{U_\gluc} du_\gluc
                \, e^{\beta_\gluc u_\gluc} e^{\beta_\ox U_\ox}                                 \\
        &-  \frac{1}{\beta_\ox}   \int_{u_\gluc^\text{min}}^{u_\gluc^a} du_\gluc
                \, e^{(\beta_\gluc - \frac{3}{7} \beta_\ox) u_\gluc} e^{\frac{3}{14} L_\maint \beta_\ox}
        -   \frac{1}{\beta_\ox}   \int_{u_\gluc^a}^{U_\gluc} du_\gluc
                \, e^{\beta_\gluc u_\gluc}
\end{split}
\intertext{which finally gives}
\begin{split}
    Z( \beta_\gluc, \beta_\ox ,\phi)    \;=\;
        &   \frac{e^{-3 K \phi \beta_\ox}}{\beta_\ox}
            \left[ \frac{e^{\beta_1 u_\gluc^b}}{\beta_1}
                    -  \frac{e^{\beta_1 u_\gluc^\text{min}}}{\beta_1} \right]
        +   \frac{e^{U_\ox \beta_\ox}}{\beta_\ox}
            \left[  \frac{e^{\beta_\gluc U_\gluc}}{\beta_\gluc}
                    - \frac{e^{\beta_\gluc u_\gluc^b}}{\beta_\gluc} \right]     \\
        &-  \frac{e^{\frac{3}{14} L_\maint \beta_\ox}}{\beta_\ox}
            \left[ \frac{e^{\beta_2 u_\gluc^a}}{\beta_2}
                    - \frac{e^{\beta_2 u_\gluc^\text{min}}}{\beta_2} \right]
        -   \frac{1}{\beta_\ox}
            \left[  \frac{e^{\beta_\gluc U_\gluc}}{\beta_\gluc}
                    - \frac{e^{\beta_\gluc u_\gluc^a}}{\beta_\gluc} \right]
    \label{eq:Z}
\end{split}
\end{align}
where
\begin{equation}
    \beta_1 \equiv \beta_\gluc + 6\beta_\ox
    \quad \text{and} \quad
    \beta_2 \equiv \beta_\gluc - \frac{3}{7}\beta_\ox.
    \label{eq:beta12}
\end{equation}



%
For evaluating the derivatives,
we first note that
\begin{equation}
    \der{}{\beta} \left(    \frac{e^{\beta^\prime u}}{\beta^\prime}     \right)
        =   \left(  u \der{\beta^\prime}{\beta}
                    + \beta^\prime \der{u}{\beta}
                    - \frac{1}{\beta^\prime} \der{\beta^\prime}{\beta}
            \right)
                \frac{e^{\beta^\prime v}}{\beta^\prime} .
\end{equation}

From \eqref{eq:beta12}, we have
\begin{equation}
    \der{\beta_1}{\beta_\gluc} = 1              \qquad;\qquad
    \der{\beta_1}{\beta_\ox} = 6              \qquad;\qquad
    \der{\beta_2}{\beta_\gluc} = 1              \qquad;\qquad
    \der{\beta_2}{\beta_\ox} = \frac{-3}{7}               .
\end{equation}


For convenience of notation, we also define
\begin{equation}
    A = -3 K \phi - \frac{1}{\beta_\ox} - \frac{6}{\beta_1}
    \qquad ; \qquad
    B = \frac{3}{14} L_\maint - \frac{1}{\beta_\ox} + \frac{3}{7} \frac{1}{\beta_2}    .
    \label{eq:AB}
\end{equation}

Using the above, we differentiate (\ref{eq:Z}) term-by-term to obtain
\begin{align}
\begin{split}
    \der{Z}{\beta_\gluc}( \beta_\gluc, \beta_\ox ,\phi)     \;=\;
        &   \frac{e^{-3 K \phi \beta_\ox}}{\beta_\ox}
            \left[ \left( u_\gluc^b - \frac{1}{\beta_1} \right)
                            \frac{e^{\beta_1 u_\gluc^b}}{\beta_1}
                    -   \left( u_\gluc^\text{min} - \frac{1}{\beta_1} \right)
                            \frac{e^{\beta_1 u_\gluc^\text{min}}}{\beta_1} \right]  \\
        &+  \frac{e^{U_\ox \beta_\ox}}{\beta_\ox}
            \left[ \left( U_\gluc - \frac{1}{\beta_\gluc} \right)
                            \frac{e^{\beta_\gluc U_\gluc}}{\beta_\gluc}
                    -   \left( u_\gluc^b - \frac{1}{\beta_\gluc} \right)
                            \frac{e^{\beta_\gluc u_\gluc^b}}{\beta_\gluc} \right]           \\
        &-  \frac{e^{\frac{3}{14} L_\maint \beta_\ox}}{\beta_\ox}
            \left[ \left( u_\gluc^a - \frac{1}{\beta_2} \right)
                            \frac{e^{\beta_2 u_\gluc^a}}{\beta_2}
                    -   \left( u_\gluc^\text{min} - \frac{1}{\beta_2} \right)
                            \frac{e^{\beta_2 u_\gluc^\text{min}}}{\beta_2} \right]  \\
        &-  \frac{1}{\beta_\ox}
            \left[ \left( U_\gluc - \frac{1}{\beta_\gluc} \right)
                            \frac{e^{\beta_\gluc U_\gluc}}{\beta_\gluc}
                    -   \left( u_\gluc^a - \frac{1}{\beta_\gluc} \right)
                            \frac{e^{\beta_\gluc u_\gluc^a}}{\beta_\gluc} \right]
    \label{eq:dZg}
\end{split}
\intertext{and}
%
\begin{split}
    \der{Z}{\beta_\ox}( \beta_\gluc, \beta_\ox ,\phi)     \;=\;
        &   \frac{e^{U_\ox \beta_\ox}}{\beta_\ox}
            \left( U_\ox - \frac{1}{\beta_\ox} \right)
                \left[ \frac{e^{\beta_\gluc U_\gluc}}{\beta_\gluc}
                        -   \frac{e^{\beta_\gluc u_\gluc^b}}{\beta_\gluc} \right]
        +   \frac{1}{\beta_\ox^2}
                \left[ \frac{e^{\beta_\gluc U_\gluc}}{\beta_\gluc}
                        -   \frac{e^{\beta_\gluc u_\gluc^a}}{\beta_\gluc} \right]           \\
        &+  \frac{e^{-3 K \phi \beta_\ox}}{\beta_\ox}
                \left[      \left( A + 6 u_\gluc^b \right)
                            \frac{e^{\beta_1 u_\gluc^b}}{\beta_1}
                        -   \left( A + 6 u_\gluc^\text{min} \vphantom{u_\gluc^b} \right)
                            \frac{e^{\beta_1 u_\gluc^\text{min}}}{\beta_1}
                \right]                                                         \\
        &-  \frac{e^{\frac{3}{14} L_\maint \beta_\ox}}{\beta_\ox}
                \left[      \left( B - \frac{3}{7} u_\gluc^a \right)
                            \frac{e^{\beta_2 u_\gluc^a}}{\beta_2}
                        -   \left( B - \frac{3}{7} u_\gluc^\text{min} \right)
                             \frac{e^{\beta_2 u_\gluc^\text{min}}}{\beta_2}
                \right] .
    \label{eq:dZo}
\end{split}
\end{align}
These analytical formulae form the basis of the self-consistency equations (\ref{eq:ulstar}) that we solve to obtain $Z_N$ and its derivatives numerically, as described in the next subsection.

\subsection{Numerical form of $Z_N(\beta_\gluc,\beta_\ox)$ and the critical line}
Using the results of the previous two subsections, we are now in a position to calculate $Z_N$ from (\ref{eq:Z_N_ident}) using $\phi^*$ and $p^*$ as functions of $\beta_\gluc$ and $\beta_\ox$, obtained numerically from the self-consistency equations (\ref{eq:pstar}) and (\ref{eq:ulstar}). Python code implementing this procedure is provided in the repository \url{https://github.com/KrishnadevN/MulticellularMetabolicNetworks}. Importantly, first- and second-order derivatives of $Z_N$ then give the expectation values and variances of net flux values, respectively, as described in Supplementary Method \ref{sec:partfunc}. These are compared with the values obtained from sampling multi-cellular flux space in Figure 3a-c from the main text. We also confirmed that numerical values for the expectation value of the net fluxes overlap nearly identically with those calculated under the mean-field measure (\ref{eq:mfmeasure}), e.g. $ \expect{u_\lact} \approx \newexpect{u_\lact}$, as might be expected in the $N \to \infty$ limit.

We remark that the analytical form of the self-consistency equations (\ref{eq:pstar}) and (\ref{eq:ulstar}) reveals some important insights into the nature of the phase-transition described by the mean-field model. Inspecting the limits of integration \eqref{eq:bounds}, we find a critical value $\phi_c = -U_\ox/3K$ such that the constraints remain constant for $\phi^*  < \phi_c$. Thus, $p^* =0$ in this region of $(\beta_\gluc,\beta_\ox)$ space, identified with the overflow phase and net lactate production as displayed in Supplementary Figure \ref{fig:pullp} and in Figure 3d from the main text. Conversely, in the regime $0>\phi^*>\phi_c$ the limits of integration become dependent on the value of $\phi^*$ such that $p^*$ steadily grows as $\phi^*$ slowly plateaus towards zero. This regime therefore corresponds to the balanced phase with minimal net lactate production. The overall effect is a critical line in phase space that delineates the separation between the balanced and overflow regimes (Supplementary Figure \ref{fig:pullp}), defined by all critical values $(\beta_\gluc^c,\beta^c_\ox)$ such that $\phi^*(\beta_\gluc^c,\beta^c_\ox) = \phi_c$. Equivalently, this can be represented as a critical value of $\newexpect{u_\lact} = -U_\ox/3K$, as displayed in Figure 4g from the main text.


\begin{figure}[htbp]
    \centering
    \includegraphics[width=\linewidth]{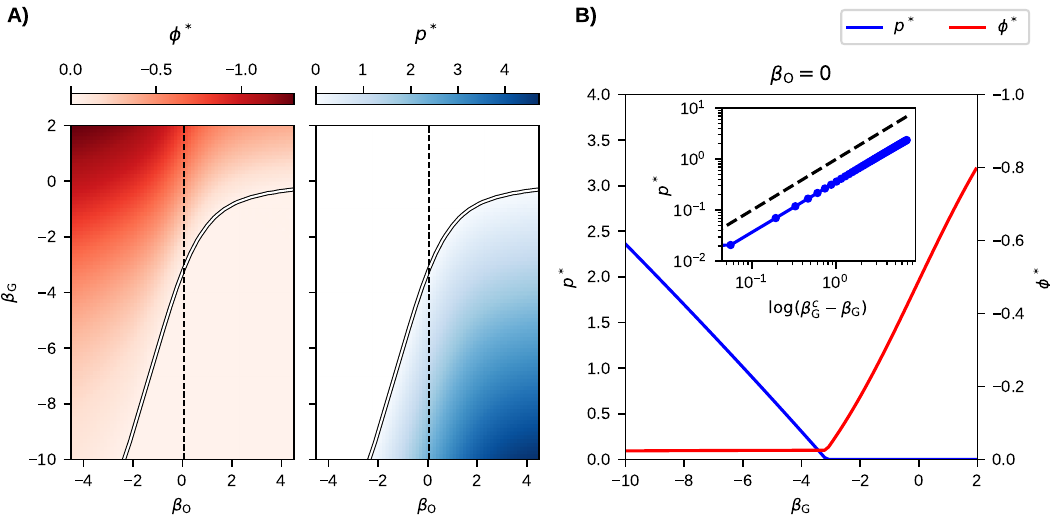}
    \caption{(A) Phase diagrams in the $(\beta_\gluc,\beta_\ox)$ plane: order parameters $\phi^*$ (left) and $p^*$ (right). The phase diagram is divided by the critical line (continuous white).
    (B) The order parameters as  a function of $\beta_\gluc$ (for $\beta_\ox=0$), corresponding to the dashed line in (A). Inset: $p^*$ as a function of $(\beta_\gluc^c - \beta_\gluc)$ in logarithmic scale showing that the critical exponent is $\alpha=1$. Here, $\beta_\gluc^c$ is the critical value of $\beta_\gluc$ corresponding to $\beta_\ox=0$.
    }
    \label{fig:pullp}
\end{figure}

\section{Background lactate term}
\label{sec:background-lactate}
\begin{figure}[h]
    \centering
\includegraphics
{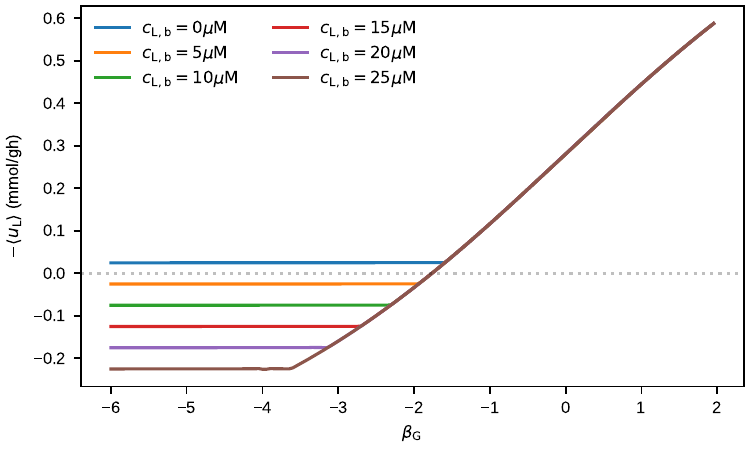}
    \caption{Plot of the mean-field lactate flux as a function of $\beta_\gluc$ at fixed $\beta_\ox=1$ for a range of relevant $c_{\lact,b}$ concentrations (0-25 $\mu$M).
    }
    \label{fig:lactrendb}
\end{figure}

\new{
While the model outlined in the previous sections forbids the net uptake of lactate (assuming none is exogenously supplied), any excess lactate produced by the culture will eventually accumulate in the growth medium to become available for net uptake at later time points.
It is straightforward to extend the model to include a background lactate concentration term (and therefore the possibility of net uptake of lactate by the culture) in the inter-cellular diffusion constraints. In presence of a background term, these read
\begin{equation}
\sum_i A_{ij} u^{(i)}_\lact \leq u_\lact^\text{max}~~~~~(\forall j)~~  \
\end{equation}
where the maximum flux limited by diffusion $u_\lact^\text{max}$ is  given by the formula
\begin{equation}
u_\lact^\text{max} = 4 \pi c_{\lact,b} D_\lact R /m
\end{equation}
with $c_{\lact,b}$ the background concentration of lactate, $D_\lact$ is the diffusion coefficient of lactate, $R$ is the cell size and $m$ is the cell mass. 
The mean-field equations in presence of a background lactate term are then modified by substituting
\begin{equation}
K\langle u_\lact \rangle \to   K\langle u_\lact \rangle - u_\lact^\text{max} ,
\end{equation}
which results in a shift of the critical point to
\begin{equation}
\phi_c =-U_\ox/3K \to  -U_\ox/3K + u_\lact^\text{max} /K
\end{equation}
and could potentially revert its sign.
We illustrate this in Fig. \ref{fig:lactrendb} showing the overflow transition for a range of relevant $c_{\lact,b}$ values and parameter values taken from Table \ref{table:parameters}.
}

\begin{figure}[h]
    \centering
    \includegraphics
    {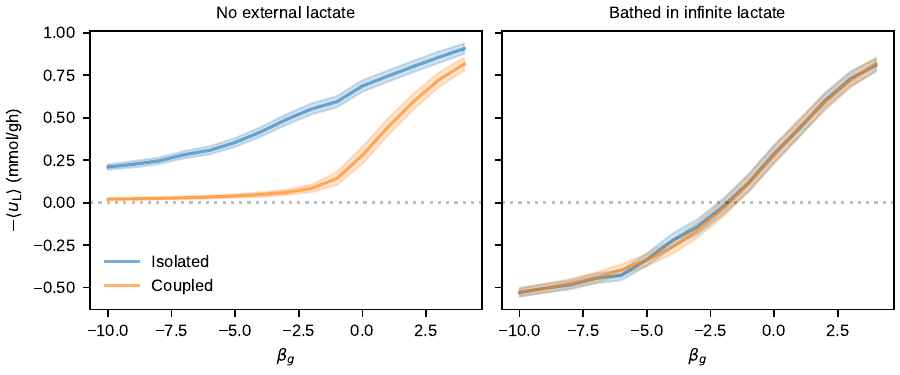}
    \caption{Numerical simulations of average lactate flux as a function of $\beta_\gluc$ at fixed $\beta_\ox=1$ for $N=150$ isolated or coupled cells. Simulations were run in the extreme cases of zero background lactate (left, $c_{\lact,b}=0$) or an infinite reservoir of lactate (right, $c_{\lact,b}\to \infty$). Shaded regions represent standard deviation on the mean.
    }
    \label{fig:lactrend2}
\end{figure}

\new{Eventually, in the limit $c_{\lact,b}\to \infty$, we find that $\phi_c$ becomes larger than the maximum uptake permitted by the oxidative capacity constraint, which causes the phase transition to disappear from the model. In this limit, the model is equivalent to a system with $N$ isolated single cells (where the inter-cellular diffusion constraints are trivially satisfied).
This is illustrated in Fig. \ref{fig:lactrend2} where we show results from simulations of the full spatial model and compare these to those from a model with isolated single cells. }

\pagebreak

\section{Limiting conditions}
\subsection{Limiting oxygen conditions}
\new{
Using the same approach for inclusion of the background lactate term, we can establish conditions where oxygen concentrations are not limiting for the critical behavior predicted by our model. Inter-cellular diffusion constraints for oxygen read
\begin{equation}
\sum_i A_{ij} u^{(i)}_\ox \leq u_\ox^\text{max}~~~~~(\forall j)~~  \
\end{equation}
where an important difference compared to the case with lactate is that oxygen can only be imported ($u^{(i)}_\ox\geq 0$). From Table \ref{table:parameters} we have
\begin{equation}
u_\ox^\text{max}/K \sim 5.75~\si{\mmolgh}   .
\end{equation}
Since this quantity represents the average local oxygen concentration experienced by single cells and is larger than the maximum oxidative capacity ($U_\ox=3~\si{\mmolgh}$), we can rationalize that intercellular diffusion constraints play little or no role for our the experimental conditions outlined in Supplementary Method \ref{sec:experiment} (in particular in regard to cellular density). We demonstrate in Supplementary Figure \ref{fig:ox} simulations of the spatial system at different background oxygen levels by plotting the average oxygen flux as a function of $\beta_\gluc$ (at fixed $\beta_\ox=1$) for $c_{O,b}\to \infty$ (ie not including the constraints), $c_{O,b}=250 \mu$M (room conditions as in experiments) and different levels of hypoxia $c_{O,b}=150,80,60 \mu$M . As expected we do not see changes for our experimental conditions while hypoxia starts to matter below $\sim 100 \mu$M.
\begin{figure}[h]
    \centering
\includegraphics[width=0.7\linewidth]{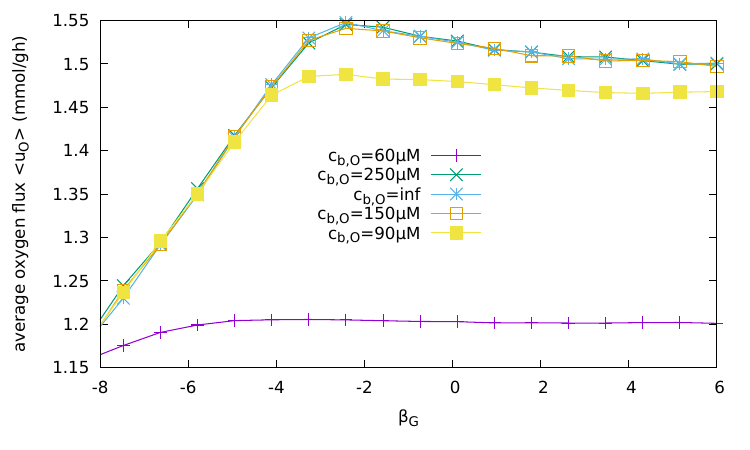}
    \caption{Average oxygen flux as a function of $\beta_\gluc$ at fixed $\beta_\ox=1$ for several $c_{O,b}$ (in $\mu$M) from simulations.
    }
    \label{fig:ox}
\end{figure}
\subsection{Limiting glucose conditions}Analogously to oxygen, the inter-cellular diffusion constraints for glucose are
\begin{equation}
\sum_i A_{ij} u^{(i)}_\gluc \leq u_\gluc^\text{max}~~~~~(\forall j)~~  \
\end{equation}
where once again glucose can only be imported ($u^{(i)}_\gluc\geq 0$). From Table \ref{table:parameters} we have
\begin{equation}
u_\gluc^\text{max}/K \sim 170 ~\si{\mmolgh}
\end{equation}
This is far higher than $U_\gluc=1 ~\si{\mmolgh}$ and we can therefore safely assume these constraints play no role in the model, provided glucose is not substantially depleted over time (a simple estimate indicate that within the 6h of experiments described in Supplementary Method \ref{sec:experiment} at most $10\%$ is consumed). We note that such an abundance of glucose is a particular feature of standard cultivation media for the cell line being analyzed.
}

\section{Inter-cellular flux correlations}
\label{sec:correlations}
\subsection{Correlations define emergent behavior in balanced phase}
\new{
In Figure 3 of the main text, we show that the mean-field model quantitatively  reproduces the trends of the average fluxes as a function of the control parameters $(\beta_\ox,\beta_\gluc)$. The mean-field approximation also highlighted that the two phases (overflow and balanced) differ by the level of cell-cell lactate flux correlations,
which become negative in the balanced phase.
This can be seen in  Supplementary Figure \ref{fig:dul} that displays, as a function of $\beta_\gluc$ (at fixed $\beta_\ox=1$) the standard deviation of  the lactate flux for a single cell, $\sigma_{u_\lact}$ (rescaled by $1/\sqrt{N}$), compared to the standard deviation of the average across cells, $\sigma_{\bar{u}_\lact}$. These two quantities are related by the formula
\begin{eqnarray}
\sigma_{\bar{u}_\lact}^2 =   \sigma_{u_\lact}^2/N + \frac{1}{N^2}\sum_{i\neq j} \langle(u_\lact^{(i)}-\langle u_\lact \rangle)(u_\lact^{(j)}-\langle u_\lact \rangle)\rangle ,
\end{eqnarray}
where the second term on the right-hand side represents the total inter-cellular correlations.
\begin{figure}[h]
    \centering
    \includegraphics[width=0.5\linewidth]{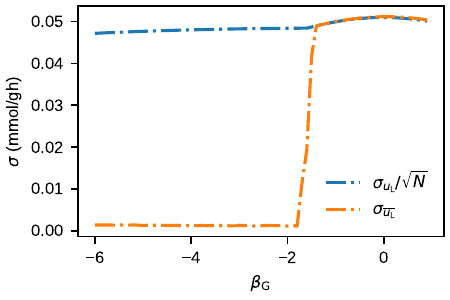}
   \caption{Standard deviation of  the lactate flux for a single cell $\sigma_{u_\lact}$, rescaled by $1/\sqrt{N}$, and  standard deviation of the mean across cells $\sigma_{\bar{u}_\lact}$ as function of $\beta_\gluc$ at fixed $\beta_\ox=1$ for mean field calculations. From mean field calculations.
    }
     \label{fig:dul}
\end{figure}
For this setup, above the point of overflow transition ($\beta_\gluc\sim-1.5$) we have that $\sigma_{\bar{u}_\lact}\sim\sigma_{u_\lact}/\sqrt{N}$, implying that correlations are zero in the overflow phase. Below it, the departure of $\sigma_{\bar{u}_\lact}$ from $\sigma_{u_\lact}/\sqrt{N}$ implies there are negative correlations among cells in the balanced phase.}

\new{
This characterizes the difference between the balanced and overflow phases, where the former is defined as a ``coordinated'' state in terms of negative inter-cellular lactate exchange flux correlations, providing a concrete definition of an emergent phenomenon (a property that single cells do not have on their own, and emerges only when they interact in a wider whole). On the other hand, by assuming cells are fully connected and equally distanced, the mean-field approximation neglects any specific spatial structure. We therefore introduce nearest neighbor correlations to study the impact of spatial orientation on coordinated behavior, as these will ultimately enable a comparison with experimental data.
}

\subsection{Nearest neighbor correlations}
\new{For $N$ cells in a given random spatial configuration (from experiments or simulations) we define average nearest neighbor inter-cellular lactate flux correlations by the expectation value of
\begin{equation}
C_\text{n.n.} = \frac{1}{N} \sum_i u_\lact^{(i)}u_\lact^{(n(i))}
\end{equation}
where $n(i)$ is the index of the first neighbor of the cell $i$, based on absolute distance. 
In the absence of pairwise correlations, the expectation value of $C_\text{n.n.}$ can be identified with the square of the average lactate flux since
\begin{equation}
    \expect{C_\text{n.n.}}_{unc}
    = \frac{1}{N} \sum_i \expect{u_\lact^{(i)} u_\lact^{(n(i))}}
    = \frac{1}{N} \sum_i \expect{u_\lact^{(i)}} \expect{u_\lact^{(n(i))}}
    = \expect{u_\lact}^2 .
\end{equation}
As shown in Supplementary Figure \ref{fig:nonlac}, left, this is reproduced by simulations of isolated single cells without exogenous lactate, as considered in Supplementary Method \ref{sec:background-lactate}.
\begin{figure}[h]
    \centering
    \includegraphics[width=0.8\linewidth]{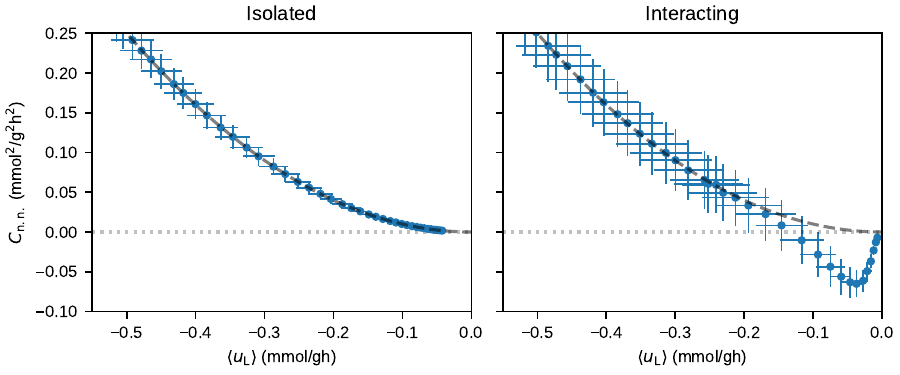}
   \caption{Left: Nearest neighbor  correlations vs  average flux for a model of isolated single cells verifies a simple scaling relation. Right: correlations vs the average flux for the coupled model violate the simple scaling relation. Simulations of the spatial model with $N=150$ cells were performed via Monte-Carlo sampling, for a sweep in $\beta_\gluc$ with $\beta_\ox=1$. Error bars denote the standard error on the quantities.
    }
     \label{fig:nonlac}
\end{figure}

On the other hand, the coupled model displays negative nearest neighbor correlations in the balanced phase, as displayed in Supplementary Figure \ref{fig:nonlac}, right.
}

\new{
\begin{figure}[h!!!!]
    \centering
    \includegraphics{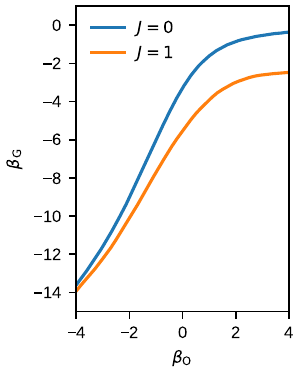}
   \caption{Critical line for the overflow threshold in the
     $(\beta_\ox,\beta_\gluc)$ plane with ($J=1$) and without ($J=0$) the phenomenological maximum entropy interaction term. Obtained numerically from simulations of the spatial model with $N=150$ cells performed via Monte-Carlo sampling on a 20$\times$20 grid.}
     \label{fig:crit}
\end{figure}
Beyond the nearest neighbor correlations that emerge as a consequence of the spatial coupling (diffusion constraints), when fitting the model to experimental data we also consider appending it as an additional term in the exponent of the Boltzmann distribution. We weight this by small phenomenological constant $J$ and augment the function $h$ in the exponent according to
\begin{equation}
h\to h+JC_\text{n.n.}   .
\end{equation}
Quadratic terms of this kind in metabolic network modeling have been studied elsewhere \cite{fernandez2020statistical} and we checked that for the value inferred in this work this term only slightly perturb the behavior of the system. We show in Fig \ref{fig:crit} the critical line in the $(\beta_\ox,\beta_\gluc)$ plane with ($J=1$) and without ($J=0$) this augmented form of the exponent, which has little effect on the critical behavior.
}

\subsection{Effects of dimensionality on inter-cellular correlations}
\new{
It is well known from statistical mechanics that mean field approximations improve upon increasing the dimension of the system, till becoming exact for many quantities of interest above a certain critical dimension\cite{parisi1998statistical}.  Thus, as we move from 2D to 3D systems we expect both nearest neighbor and total correlations correlations to become better approximated using the mean-field model.
This hypothesis is verified in Supplementary Figure \ref{fig:dim}, where for both types of correlations 3D cultures are found to lie between those from 2D cultures and the mean-field model. Here the total density $\rho_{3d}$ of the 3D system was chosen to match the magnitude of left-hand side of the inter-cellular diffusion constraint for 2D cultures by imposing
\begin{equation}
\rho_{2d} = \rho_{3d} L
\end{equation}
where $L=1cm$ is the total system size.
By comparison with nearest neighbor correlations, which are most pronounced for 2D cultures, total correlations for both 2D and 3D cultures do not reach levels observed for the mean-field model (where both nearest neighbor and total correlations are the same by construction), which explains the small discrepancy between predictions for $\sigma_{\bar{u}_\lact}$ in the balanced phase from the mean-field model versus simulations reported in Figure 3 of the main text.
}
\begin{figure}[h]
    \centering
\includegraphics[width=0.48\linewidth]{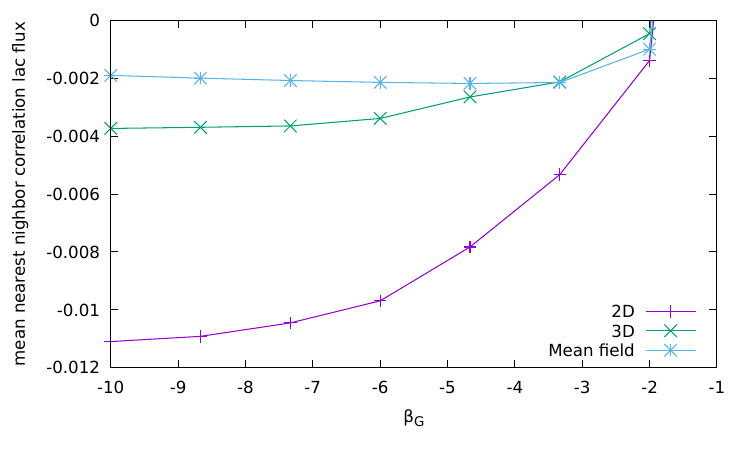}
\includegraphics[width=0.48\linewidth]{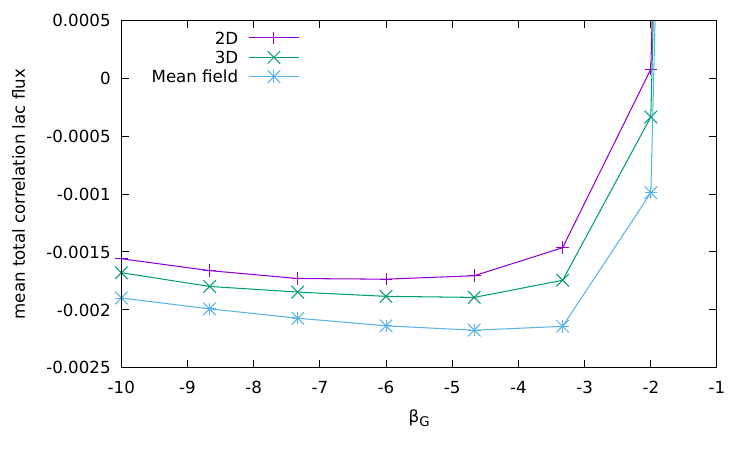}
    \caption{Average lactate flux nearest neighbor (left) and total (right) correlations as a function of $\beta_\gluc$ at fixed $\beta_\ox=1$ for systems of different dimensionality (2D, 3D and fully connected mean-field approximation).
    }
    \label{fig:dim}
\end{figure}

\section{The experimental dataset}
\label{sec:experiment}

This work includes experimental data extracted from \cite{onesto2022probing}. Briefly, pH-sensing hybrid nanofibers were fabricated through electrospinning technique starting from polycaprolactone (PCL)
polymer dissolved at a concentration of 10\% (w/v) in chloroform 
and dimethyl sulfoxide (DMSO).
The PCL solution was mixed with \qty{36}{\mg\per\mL} of ratiometric pH-sensing microparticles based on the pH-indicator dye fluorescein 5(6)-isothiocyanate (FITC) 
and the reference dye rhodamine B isothiocyanate (RBITC) 
synthesized as described previously \cite{chandra2022}.
For cell culture experiments, the pH-sensing nanofibers were 
placed into a $\mu$-slide 4 well chamber slide 
with \num{4e4} cells/well of human pancreatic cancer cell line AsPC-1 
and cancer-associated fibroblasts (CAFs)
in the ratio 70\% CAFs and 30\% AsPC-1 tumor cells at \qty{37}{\degreeCelsius} in a humidified 5\% \ch{CO2} incubator. 
The fluorescence response of the pH-sensing nanofibers during cell culture was monitored via time-lapse confocal laser scanning microscopy 
for 6 hours with a time interval of 10 minutes, maintaining a temperature of \qty{37}{\degreeCelsius}. 
Before the experiment, \qty{200}{\ul} of Leibovitz L15 medium 
were added to the samples and the calibration of the whole system (pancreatic tumor and pancreatic stromal cells seeded on the pH-sensing nanofibers) was performed. 
Single-cell fermentation fluxes were inferred by inverse modeling the  scalar pH field under the approximation of steady state diffusion from spherical sources and sinks. For further details see\cite{onesto2022probing}.

\subsection{Flux correction due weak acid reversible proton binding}
Experimental fluxes from the dataset \cite{onesto2022probing} have been re-scaled here by a factor $7$, which is approximately the ratio between the diffusion constant of protons and lactate, respectively (see Supplementary Table 1). The output of experiments in \cite{onesto2022probing} in fact gives more precisely the ratio between the cellular uptake flux and the diffusion constant, the latter assumed to be the one of protons. This assumption, equivalent to a complete dissociation of the acid is incorrect for the case of a weak acid like lactate we are considering here.

\subsection{Conversion of lactic acid level to pH}

We will examine the simple case of a uniform dilute solution of acid.
The situation can be treated as if the acid \ce{HA} dissolves first in molecular form and ionises until equilibrium is reached. Let this initial concentration of acid be $c$.
The two contributions to \ce{H3O^+} in solution are (i) the ionisation of \ce{HA} and (ii) the self-ionisation of water. Let these concentrations be $h_a$ and $h_w$ respectively.
We have,
\begin{align*}
    \ce{ [H3O^+]    &=  [A^-] + [OH^-] } = h_a + h_w ,   \\
    \ce{[HA]}       &=  c - h_a .
\end{align*}

Using the expressions for $\K{a}$ and $\K{w}$, we can write
\begin{alignat}{3}
    \K{a}  &=  \ce{\frac{[H3O^+][A^-]}{[HA]}} &&= \frac{(h_a + h_w)h_a}{c - h_a} ,
    \label{eq:Ka}   \\
    \K{w}  &=  \ce{ [H3O^+][OH^-] }           &&= (h_a + h_w)h_w    .
    \label{eq:Kw}
\end{alignat}

We need to solve the pair of equations for $h_a$ and $h_w$, given $\K{a}$, $\K{w}$ and $c$. It is convenient to substitute for $(h_a + h_w)$ and $h_w$ in \eqref{eq:Kw} using \eqref{eq:Ka}. Substituting
\begin{equation}
    (h_a + h_w)  =  \frac{\K{a} (c - h_a)}{h_a}
    \quad\text{and}\quad
    h_w  =  \frac{\K{a} (c - h_a)}{h_a} - h_a
    \label{eq:hahw}
\end{equation}
into \eqref{eq:Kw} gives
\begin{equation}
    \K{w}   =   \frac{\K{a} (c - h_a)}{h_a}
                \left[  \frac{\K{a} (c - h_a)}{h_a} - h_a  \right] ,
\end{equation}
which can be rearranged to get an implicit equation for $h_a$
\begin{equation}
    h_a^2  =  \frac{\K{a}^2 (c - h_a)^2} {\K{w} + \K{a} (c - h_a)} .
    \label{eq:ha}
\end{equation}

This equation can be solved using successive approximations, first assuming $h_a \ll c$, which is valid for a weak acid with low degree of ionisation, to obtain
\begin{equation}
    h_a     \approx     \frac{\K{a} c} {\sqrt{\K{w} + \K{a} c}}
    \label{eq:ha_approx}
\end{equation}
and substituting the value back in \eqref{eq:ha} for an improved estimate. The approximation $c-h_a \approx c$ works best when $c/\K{a}>100$ or greater.
Further, in the case of lactic acid, $\K{a} \approx \num{1.38e-4}$ and $\K{w} \approx \num{e-14}$ so that, unless the acid solution is very dilute (i.\,e.\ if $\K{a} (c - h_a) \approx \K{a} c \gg \K{w}$), \eqref{eq:ha} can be approximated as
\begin{equation}
    h_a^2   \approx     \K{a} (c - h_a)
    \quad\text{or}\quad
    h_a     \approx     \sqrt{\K{a} c}  .
\end{equation}

To obtain a first approximation of \ce{[H3O^+]} and hence the pH of the medium, \eqref{eq:ha_approx} could be substituted in \eqref{eq:hahw} to obtain
\begin{equation}
    \ce{[H3O^+]}  =  (h_a + h_w)    =          \frac{\K{a}c}{h_a} - \K{a} \approx    \sqrt{\K{w} + \K{a} c} - \K{a} .
\end{equation}

It is to be noted that \eqref{eq:hahw} cannot be used directly when $c \to 0$, as both the numerator and denominator tend to 0. Instead, when $h_a=0$, $\ce{[H3O^+]} = h_w = \sqrt{\K{w}}$ from \eqref{eq:Kw}. It is possible to solve \eqref{eq:ha} numerically and substitute the value in \eqref{eq:hahw} to obtain \ce{[H3O^+]} as a function of $c$.

\begin{figure}[h]
    \centering
    \includegraphics[width=0.75\linewidth]{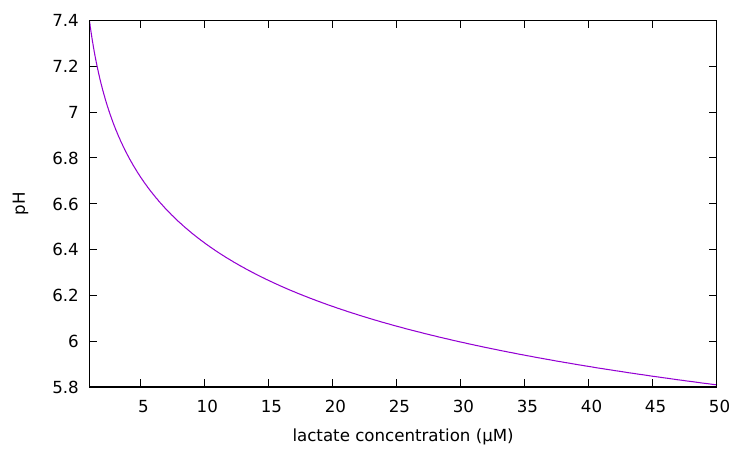}
    \caption{pH as a function of the lactate concentration.
    }
\end{figure}

In our experimental setup, we do not have direct measurements of the concentration of acids. Instead, we have measurements of pH and hence of \ce{[H3O^+]}.
So in \eqref{eq:Ka} and \eqref{eq:Kw}, the known quantity is $h_a + h_w$ and the unknown is $c$. We seek an expression for $c$ in terms of $h = h_a + h_w$, given \K{a} and \K{w}.
We can rewrite \eqref{eq:Ka} and \eqref{eq:Kw} in terms of $h$ and substitute for $h_a$ in \eqref{eq:Ka} using \eqref{eq:Kw} as
\begin{alignat}{3}
    \K{w}  = h h_w  =  h (h - h_a)
        &\implies&  h_a &=  \left( h - \frac{\K{w}}{h} \right)                  \\
    \K{a}  = \frac{h h_a}{c - h_a}
        &\implies&  \frac{h}{\K{a}}  &=  \frac{c}{h_a} - 1          \nonumber   \\
        &\implies&  c   &=  h_a \left( 1 + \frac{h}{\K{a}} \right)  \nonumber   \\
        &\implies&  c   &=  \left( h - \frac{\K{w}}{h} \right)
                            \left( 1 + \frac{h}{\K{a}} \right)      \label{eq:lactconc}
\end{alignat}
It is easily verified that when $h = h_w = \sqrt{\K{w}}$, \eqref{eq:lactconc} correctly gives $c=0$.
This function $c(h)$ can be numerically inverted to obtain a lookup table for the inverse function $h(c)$, from which we calculate pH as a function of lactate concentration in the medium.
We further add an offset term to this formula to match the experimental pH, phenomenologically modeling the buffering of the medium. The resulting conversion curve is shown in Supplementary Figure 8.


\subsection{Background lactate accumulation}
\label{sec:lactate-accumulation}
\begin{figure}[h]
    \centering
    \includegraphics[width=0.75\linewidth]{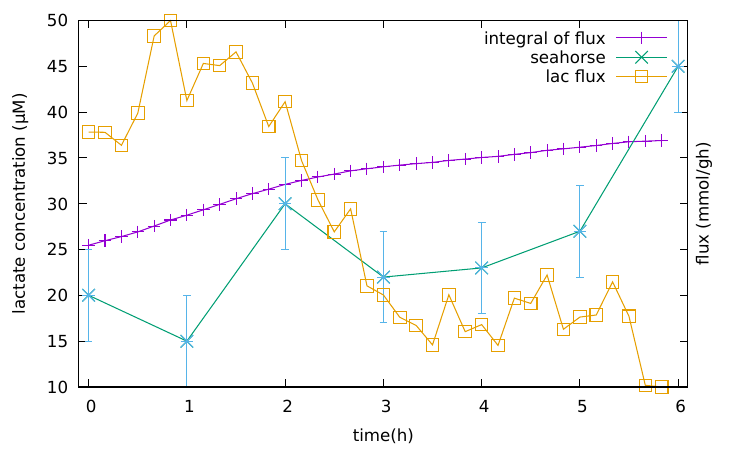}
    \caption{Experimental lactate concentration as function of time, obtained from a lactate dehydrogenase assay (stars) and by integrating over average cell lactate flux (crosses). Also shown is the average lactate flux (squares). Data from \cite{onesto2022probing}
    }
    \label{fig:lactrend}
\end{figure}

\new{As described previously, although no exogenous source of lactate is supplied to the growth medium, lactate produced by cells can accumulate to become available for net uptake at later time points. Supplementary Figure \ref{fig:lactrend} displays the experimental time course of total lactate concentrations as measured independently using a Seahorse assay with lactate dehydrogenase (see \cite{onesto2022probing} for further details) and compared with the integral of the net lactate flux. Both measurements confirm that background concentrations of lactate are increasing in time as it is being produced by cell cultures, which motivates inclusion of the background lactate term described in Supplementary Method \ref{sec:background-lactate} when fitting the model to data. The implementation of the background term in the inverse modeling of experimental data has been performed by matching the experimentally measured lactate level with the sum of the of the average lactate level in the observed frame due to flux from  cells and the boundary constant term according to
\begin{equation}
c_{\lact,exp}=\langle c_\lact \rangle + c_{\lact,b}  .
\end{equation}}

\pagebreak

\section{Inverse modeling  experimental data}

\subsection{Likelihood function}
\label{sec:likelihood}
The maximum entropy model we are considering is determined by two parameters, which are the Lagrange multipliers $\beta_\gluc$ and $\beta_\ox$ that fix the average net glucose and oxygen fluxes, respectively, across the cell population. Equivalently, one could consider any other linearly independent combinations of these Lagrange multipliers, such as those fixing the average net fluxes for ATP and lactate production. As we remarked at the end of Supplementary Method \ref{sec:partfunc}, the setting could be modified beyond the homogeneity assumption and allowing for $2N$ parameters to be inferred. This would make the problem computationally more difficult but still feasible with the help of expectation propagation techniques \cite{braunstein2017analytic,pereiro2022inference,muntoni2022relationship}.

Ideally, experimental data should provide information about independent sets of net metabolic fluxes. In this respect, one caveat of our experimental data is that we have access only to the set of single cell lactate fluxes at each experimental time point. However, we can instead exploit the single-cell resolution nature of these data, to fit the  moments of the observed single-cell experimental lactate flux distribution, which are predicted as functions of $\beta_\gluc$ and $\beta_\ox$ in our model.
This leads to a predictive model for the observables: average net lactate flux, fluctuations (as measured by the standard deviation) and \new{nearest neighbor correlations}, which can be fitted by \new{maximizing the likelihood of the experimental data. Assuming that sources of noise are Gaussian and independently distributed, the log-likelihood $\mathcal{L}_\gamma(\vec{\beta}_\ox,\vec{\beta}_\gluc)$ can be written as the sum of two terms. The first is the  residue and/or distance between the experimental observables and the corresponding predictions the model, while the second is a regularizer term that prevents over-fitting by ensuing continuity of the parameters across adjacent time points, controlled by the hyper-parameter $\gamma$:
\begin{multline}
    \mathcal{L}_\gamma(\vec{\beta}_\ox,\vec{\beta}_\gluc)
    = \sum_t - \bigg[
                    \frac{(\avgul_{\text{exp},t} - \expect{u_\lact}_{\text{mod},t})^2}
                        {2 \sigma^2_{\text{mean},t}}
                +   \frac{(\sigma_{u_\lact,\text{exp},t} - \expect{\sigma_{u_\lact}}_{\text{mod},t})^2}
                        {2 \sigma^2_{\text{fluc},t}}
                +   \frac{(C_{\text{n.n.,exp},t} - \expect{C_\text{n.n.}}_{\text{mod},t})^2}
                        {2\sigma^2_{\text{corr},t}}                 \\
                - \gamma \left[ (   \beta_\ox(t)-\beta_\ox(t+1))^2
                                +  (\beta_\gluc(t)-\beta_\gluc(t+1))^2 \right]
                \bigg] .
\end{multline}

Here the sum is taken over experimental time points $t=0, 1,...,T$ (with $\vec{\beta}_\ox \equiv (\beta_\ox(0),\beta_\ox(1), \beta_\ox(2),..., \beta_\ox(T) )$ and same for $\vec{\beta}_\gluc$) and experimental errors were estimated via jackknife resampling. The most probable values of the parameters are retrieved by maximizing $\mathcal{L}$ as described in the next subsection, while errors on the inferred parameters can be estimated by considering that  knowledge of $\mathcal{L}$ leads to their posterior probability distribution approximated (up to to a constant) by
\begin{equation}
\text{prob}(\vec{\beta}_\ox,\vec{\beta}_\gluc) \propto e^{\mathcal{L}}   .
\end{equation}

\subsection{The sampling method}
The parameters have been sampled according to the posterior probability distribution via a Metropolis Monte-Carlo algorithm according to the following steps:

\begin{enumerate}
\setcounter{enumi}{-1}
\item Initial data: a set of grids in the $(\beta_\ox,\beta_\gluc)$ plane that have been calculated from the model with the values of the observables of interest (average, single cell fluctuations and nearest neighbor correlations of the lactate flux), one for each experimental time point, with increasing level of background lactate. This is done using the method described in Supplementary Method \ref{sec:sampling}. Define the starting value for the Markov chain $(\vec{\beta}_\ox^n,\vec{\beta}_\gluc^n)$, $n=0,1,2,\dots$ with $(\vec{\beta}_\ox^{0},\vec{\beta}_\gluc^{0}) = (\vec{0},\vec{0})$.
\item Choose uniformly at random an experimental time point $t^* \in [0,T]$ and uniformly at random from the grid corresponding to that time point a value $(\beta_\ox^*,\beta_\gluc^*)$. Construct a proposal by substituting in the current vector the component  corresponding to that time point:
\begin{equation*}
    (\vec{\beta}_\ox^{\text{prop}},\vec{\beta}_\gluc^{\text{prop}})=  ((\beta_\ox^n(t=0),\beta_\gluc^n(t=0)),\dots,(\beta_\ox^*(t^*),\beta_\gluc^*(t^*)),\dots,(\beta_\ox^n(t=T),\beta_\gluc^n(t=T))) .
\end{equation*}
\item Calculate the difference $\Delta \mathcal{L} = \mathcal{L}(\vec{\beta}_\ox^{\text{prop}},\vec{\beta}_\gluc^{\text{prop}}) - \mathcal{L}(\vec{\beta}_\ox^{n},\vec{\beta}_\gluc^{n})$.
\item (Metropolis) Accept the proposal if $\Delta \mathcal{L}\geq0$ or, if  $\Delta \mathcal{L}<0$, extract a random number $r \in [0,1]$ at uniform and accept the proposal if $r<e^{\Delta \mathcal{L}}$. If proposal accepted go to 4, otherwise go to 5.
\item Set
\begin{equation*}
(\vec{\beta}_\ox^{n+1},\vec{\beta}_\gluc^{n+1}) = (\vec{\beta}_\ox^{\text{prop}},\vec{\beta}_\gluc^{\text{prop}})
\end{equation*}
and go to 1.
\item Set
\begin{equation*}
(\vec{\beta}_\ox^{n+1},\vec{\beta}_\gluc^{n+1}) = (\vec{\beta}_\ox^{n},\vec{\beta}_\gluc^{n})
\end{equation*}
and go to 1.
\end{enumerate}

After disregarding an initial transient (around $10^5$ points), we collected $\sim10^6$ samples from which we evaluated the statistical quantities of interest.

\subsection{Training and test set split: setting the hyper-parameter $\gamma$}

We tested the predictive capabilities of the model with  standard methods of statistical inference.
We split the $M=36\times 3=118$ observables in two sets: a training set, comprising $90\%$ of the observables taken uniformly at random, over which we perform the fit (i.e., values of the observables are included in the definition of $\mathcal{L}$), and a test set with the remaining $10\%$ values used to evaluate model predictions. This procedure was repeated $10^4$ times for different realization of the random split and it has been evaluated as a function of the hyper-parameter.
The reduced $\chi^2$ has been calculated with respect to the training and test set as a function of the hyper-parameter $\gamma$. The trend is monotonously increasing for the training set (starting from $\chi^2=$ at $\gamma=0$) while for the test set it decreases sharply up to $\gamma=1$ where it assumes approximately the value of the of the training set (see fig \ref{fig:hyper}). The trade-off is thus optimal  around $\gamma=2$, the value we finally set.

\begin{figure}[h!]
    \centering
    \includegraphics[width=0.6\linewidth]{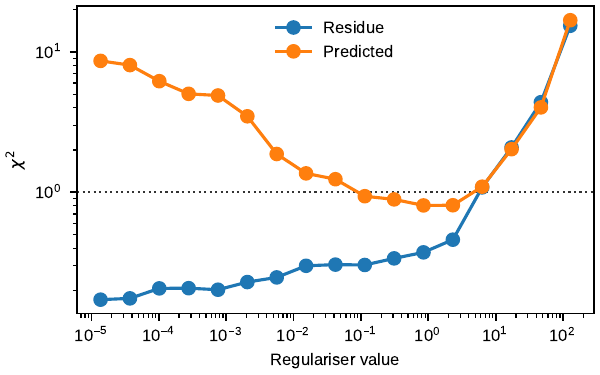}
\caption{Reduced $\chi^2$ (average square residue) for the fitted and predicted data as a function of $\gamma$}
\label{fig:hyper}
\end{figure}
}

\subsection{Fitting the model without correlations}
\new{
In the first instance, we trialed fitting the two model parameters $\beta_\gluc, \beta_\ox$ to two observables, and chose the average and standard deviation (fluctuations) of lactate flux $\langle u_\lact \rangle$ and $\sigma_\lact$, respectively, by excluding the nearest neighbor correlations term from the likelihood function. In this scenario, we see by comparing the grids of model predictions (Supplementary Figure \ref{fig:noninterfit2}) that a model with isolated single cells (non-interacting) will never be able to reproduce these experimental data. On the other hand, the model involving inter-cellular diffusion constraints (interacting model) is successful in producing a set of $\langle u_\lact \rangle$ and $\sigma_\lact$ values that match the experimental observations (compare Supplementary Figure \ref{fig:fitrestrict} with \ref{fig:noninterfit}). However, the nearest neighbor correlations predicted by the interacting model fail to reproduce those observed experimentally (Supplementary Figure \ref{fig:fitrestrict}) and we find inferred ($\beta_\gluc, \beta_\ox$) points have large errors bars extending deeply into the balanced phase that seem to self-organize around the critical lines (Supplementary Figure \ref{fig:phdiagrestrict}). The latter seems to be an indication of over-fitting since similar biases has been noted in inverse modeling neural systems.

\begin{figure}[h!]
    \centering
    \includegraphics[width=\linewidth]{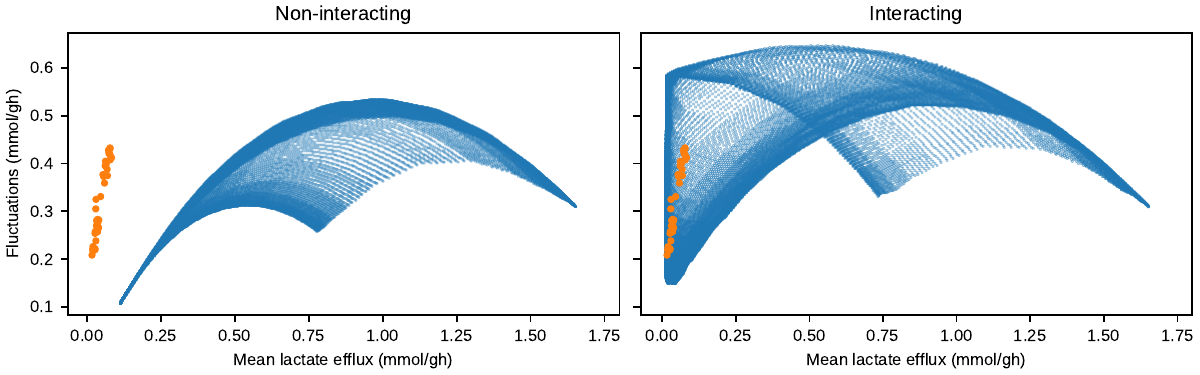}
\caption{Grid of model predictions (blue) compared to experimental data (orange) for models with isolated single cells (left, non-interacting model) and cells coupled by inter-cellular diffusion constraints (right, interacting model).}
\label{fig:noninterfit2}
\end{figure}

\begin{figure}[h!]
    \centering
    \includegraphics[width=\linewidth]{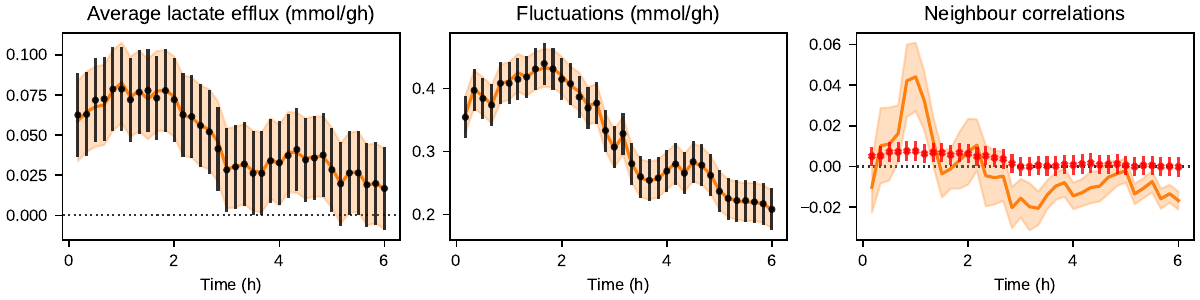}
   \caption{Fit of the interacting model (black dots) to averages and fluctuations of the experimental lactate flux (orange lines). Nearest neighbor correlations (red) were excluded from the fit. See Supplementary Method~\ref{sec:likelihood} for details on error estimation.}
   \label{fig:fitrestrict}
\end{figure}

\begin{figure}[h!]
    \centering
    \includegraphics[width=\linewidth]{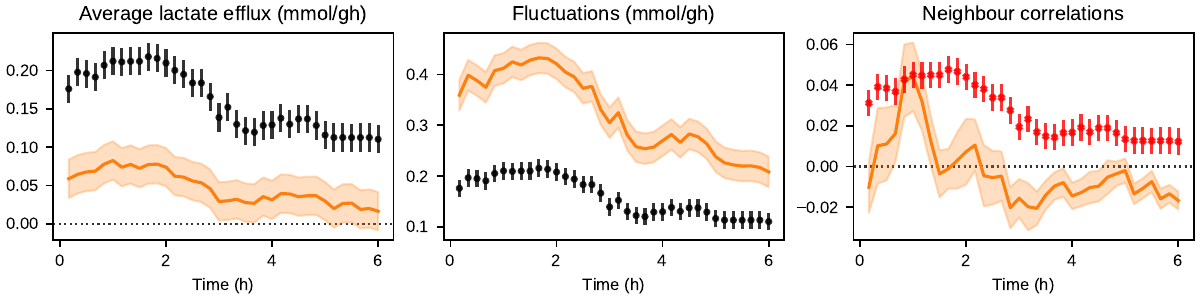}
\caption{Fit of the non-interacting model (black dots) to averages and fluctuations  of the experimental lactate flux (orange lines). Nearest neighbor correlations (red) were excluded from the fit. See Supplementary Method~\ref{sec:likelihood} for details on error estimation.}
\label{fig:noninterfit}
\end{figure}

\begin{figure}[hp!]
    \centering
    \includegraphics[width=0.4\linewidth]{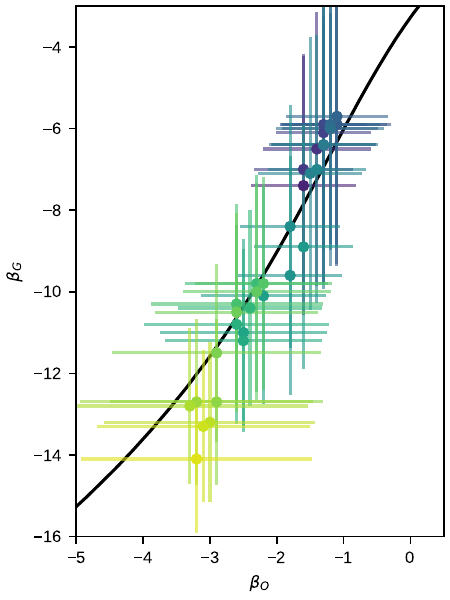}
   \caption{Inferred points in the $(\beta_\ox,\beta_\gluc)$ phase space obtained from fitting the model to averages and standard deviations of the lactate flux, excluding correlations. The critical line obtained from the mean-field model without exogenous lactate is shown in black.}
   \label{fig:phdiagrestrict}
\end{figure}

\subsection{Fitting the model with correlations}

In an effort to improve the fit of the interacting model, we thus included the term with nearest neighbor correlations in the likelihood. We note that a non-interacting model with isolated single cells cannot reproduce these correlations by construction. In the absence of a background lactate term, the interacting model then presented a reasonable fit of these experimental data, reproducing the negative nearest neighbor correlations observed in the balanced phase (Supplementary Figure \ref{fig:nolac}).

\begin{figure}[hp!]
    \centering
    \includegraphics[width=\linewidth]{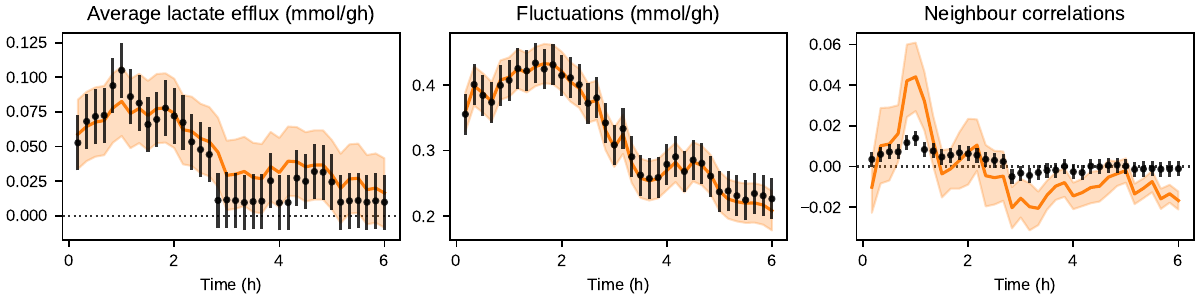}
   \caption{Fit of the interacting model (black dots) to the average, fluctuations and nearest neighbor correlations of the experimental lactate flux (orange lines), neglecting background lactate accumulation. See Supplementary Method~\ref{sec:likelihood} for details on error estimation.}
   \label{fig:nolac}
\end{figure}

However, since the background concentration of lactate is both predicted and observed experimentally to increase over time, we next assessed a fit of the same data with the background lactate term estimated as described previously (Supplementary Methods \ref{sec:background-lactate} and \ref{sec:lactate-accumulation}).
This lead to a mutual inconsistency between the three observables predicted by the model. As shown in Supplementary Figure \ref{fig:fitnoJ}, based on diffusion constraints alone, the interacting model cannot explain the negative nearest neighbor correlations observed experimentally when the background concentrations of lactate are high. This motivated an inclusion of an additional phenomenological term in the Boltzmann exponent as discussed in Supplementary Method \ref{sec:correlations} and main text. The final value of this phenomenological term has been fixed upon performing the inference for a range of six different values (Supplementary Figure \ref{fig:fitJ}).

\begin{figure}[hp!]
    \centering
    \includegraphics[width=\linewidth]{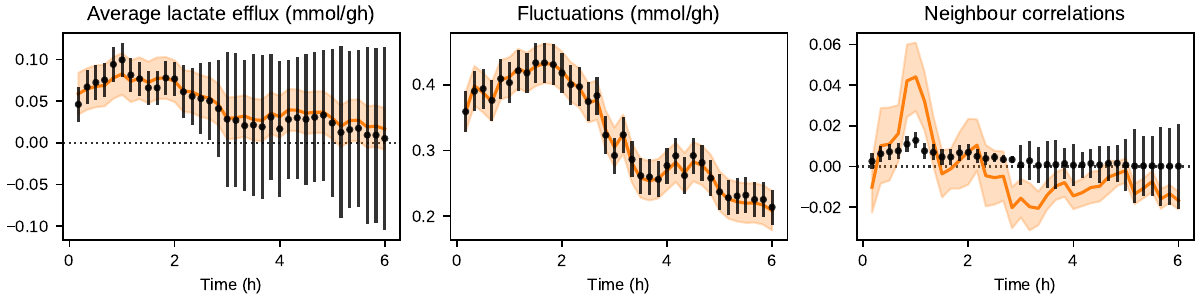}
   \caption{Fit of the interacting model (black dots) to the average, fluctuations and nearest neighbor correlations of the experimental lactate flux (orange lines), including background lactate accumulation. See Supplementary Method~\ref{sec:likelihood} for details on error estimation.}
   \label{fig:fitnoJ}
\end{figure}

\begin{figure}[h!]
    \centering
    \includegraphics[width=0.5\linewidth]{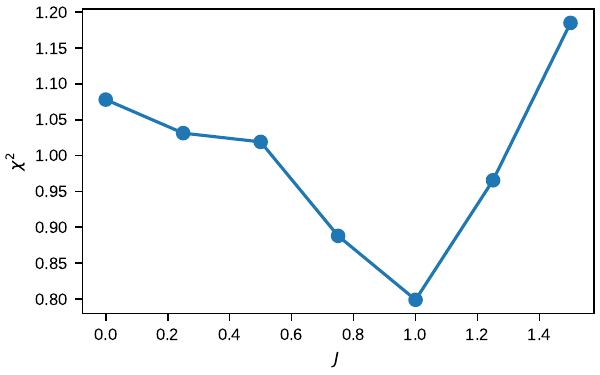}
   \caption{Reduced $\chi^2$ as  function of $J$}
   \label{fig:fitJ}
\end{figure}
}

\pagebreak

\section{Modeling single-cell oxygen dynamics}

From the inverse modeling, we  have an estimate of the average oxygen flux
$\expect{u_\ox}_\text{ME}$ that we can complement with the single cell measured value $u_\lact^{(n)}$ of the lactate flux to obtain single cell estimates of the oxygen fluxes,
once again, using the maximum entropy method.
The lower bound $\mathcal{L}_\ox^{(n)}$ and upper bound $\mathcal{U}_\ox^{(n)}$ for oxygen flux $u_\ox^{(n)}$ is calculated using the measured lactate flux $u_\lact^{(n)}$ as
\begin{equation}
\begin{aligned}
    \mathcal{L}_\ox^{(n)}   &=  \max \bigg( 0,\; 3\,u_\lact^{(n)},\;
                                            \frac{M+u_\lact^{(n)}}{5}   \bigg)      \\
    \mathcal{U}_\ox^{(n)}   &=  \min \Big( U_\ox,\; 6\,U_\gluc + 3\,u_\lact^{(n)} \Big)   .
\end{aligned}
\end{equation}
For cells with $u_\lact^{(n)}<-2\,U_\gluc$, we enforce $u_\ox^{(n)}=0$.
We then enforce a maximum entropy constraint on the average oxygen flux with a Lagrange multiplier $\xi$ that is independent of the cell index.
To obtain the average oxygen flux for the cell $n$ we integrate the marginal Boltzmann probability distribution over the segment $[ \mathcal{L}_\ox^{(n)}\mathcal{U}_\ox^{(n)}]$
\begin{equation}
 \big\langle u_\ox^{(n)} \, \big\rangle = \frac{\int_{\mathcal{L}_\ox^{(n)}}^{\mathcal{U}_\ox^{(n)}} u e^{\xi u} \, du}{\int_{\mathcal{L}_\ox^{(n)}}^{\mathcal{U}_\ox^{(n)}} e^{\xi u} \, du}
\end{equation}
that gives

\begin{equation}
    \big\langle u_\ox^{(n)} \, \big\rangle   =
        \frac   { \mathcal{U}_\ox^{(n)} e^{\xi \, \mathcal{U}_\ox^{(n)}}
                    -   \mathcal{L}_\ox^{(n)} e^{\xi \mathcal{L}_\ox^{(n)}} }
                { e^{\xi \, \mathcal{U}_\ox^{(n)}} - e^{\xi \mathcal{L}_\ox^{(n)}} }
    -   \frac{\hspace{0.2em} 1 \hspace{0.2em}}{\xi}  .
    \label{eq:oxeq-1}
\end{equation}

We then match the average $\big\langle u_\ox^{(n)} \, \big\rangle$ with $\expect{u_\ox}_\text{ME}$
\begin{equation}
    \frac{1}{N}     \sum_{i=1}^N
    \big\langle u_\ox^{(n)} \, \big\rangle
    =   \expect{u_\ox}_\text{ME}
\end{equation}
which gives
\begin{equation}
    \frac{\hspace{0.2em} 1 \hspace{0.2em}}{\xi}
    =
    \frac{1}{N}     \sum_{n=1}^N    \;
        \frac   { \mathcal{U}_\ox^{(n)} e^{\xi \, \mathcal{U}_\ox^{(n)}}
                    -   \mathcal{L}_\ox^{(n)} e^{\xi \mathcal{L}_\ox^{(n)}} }
                { e^{\xi \, \mathcal{U}_\ox^{(n)}} - e^{\xi \mathcal{L}_\ox^{(n)}} }
    -   \expect{u_\ox}_\text{ME}
    \label{eq:oxeq-2}
\end{equation}
The value of $\xi$ obtained solving numerically equation \eqref{eq:oxeq-2} can then be 
used to estimate $ \big\langle u_\ox^{(n)} \, \big\rangle$ from \eqref{eq:oxeq-1}.

This procedure, with our data, does give  large errors, i.e.\ 
there are many equivalent models with inferred single cell oxygen flux reproducing the lactate data that are compatible with the constraints. Our inferred model can be thus regarded as a ``sloppy'' one in the sense of \cite{gutenkunst2007universally} and does not lead to  real single-cell measurements of the oxygen flux. This is a task that could be finally be achieved with the help of nanometric sensing  \cite{grasso2024highly}.


\bibliographystyle{plain}
\bibliography{supplementary}%